\documentclass[12pt]{article}
\usepackage{custom}

\usepackage{xr}
\begin{document}
\title{Broad Validity of the First-Order Approach in Moral Hazard%
\thanks{We are especially grateful to Lucas Maestri for originating key ideas in this paper. We thank Daniel Gottlieb, Rongzhu Ke, Humberto Moreira, Marcus Opp, Juuso Toikka and seminar participants for valuable comments. Replication code is available at \url{https://github.com/eduardomazevedo/azevedo-wolff-2025}, and supplementary information is available at \url{https://eduardomazevedo.github.io/papers/azevedo-wolff-foa-supplementary-material.pdf}.}
}

\author{Eduardo M. Azevedo% 
\thanks{Wharton: 3733 Spruce Street, Philadelphia, PA 19104: eazevedo@wharton.upenn.edu, \url{https://eduardomazevedo.github.io/.}}
\and Ilan Wolff% 
\thanks{Wharton: 3733 Spruce Street, Philadelphia, PA 19104. ijwolff@wharton.upenn.edu.}
}

\date{
First version: March 7, 2025 \\
This version: \today}

\maketitle

\begin{abstract}%
We consider the standard moral hazard problem with limited liability. The first-order approach (FOA) is the main tool for its solution, but existing sufficient conditions for its validity are restrictive.
Our main result shows that the FOA is broadly valid, as long as the agent's reservation utility is sufficiently high.
In basic examples, the FOA is valid for almost any positive reservation wage.

We establish existence and uniqueness of the optimal contract. We derive closed-form solutions with various functional forms. We show that optimal contracts are either linear or piecewise linear option contracts with log utility and output distributions in an exponential family with a linear sufficient statistic (including Gaussian, exponential, binomial, geometric, and gamma distributions). We provide an algorithm that finds the optimal contracts at negligible computational cost, whether or not the FOA is valid.
\end{abstract}

\newpage

\section{Introduction}
\label{sec:introduction}
The principal-agent problem with moral hazard is as follows. The principal hires the agent to take an action $a$ in $\mathbb R _ +$ that affects the distribution $f(y|a)$ of output $y$. The principal can only condition payments on realized output, and designs a contract $w(y) \geq 0$ to provide incentives to the agent. The agent chooses the action $a$ to maximize her utility from wages minus her cost of effort $c(a)$, and the principal chooses the contract to maximize expected profits.

The main solution method for this problem is the first-order approach, which assumes that only local deviations in $a$ are binding, and yields a simple formula for the optimal contract.%
\footnote{%
See \textcite{holmstrom1978incentives}, the excellent survey \textcite{georgiadis2022contracting}, and recent work in \textcite{conlon2009two,kadan2017existence,chade2020no,chaigneau2022should,castro2024disentangling}.
}
Because of tractability, a vast literature simply assumes that the first-order approach is valid, and many applied papers make restrictive assumptions to avoid the issue of nonlocal deviations.%
\footnote{%
\textcite{mirrlees1999} showed that the first-order approach is not always valid. Examples of papers assuming the first-order approach include \textcite{jewitt2008moral, moroni2014existence, chaigneau2022should,castro2024disentangling}. Virtually all early work, including the seminal papers by \textcite{holmstrom1978incentives, holmstrom1979moral}, and \textcite{zeckhauser1970medical} assumes it. \textcite{holmstrom1979moral} notes that 
``one has to assume that the agent's optimal choice of action is unique
for the optimal [...] This assumption seems very difficult to validate [...] and regrettably we have to leave the question about its validity open.'' In applied work, restrictive assumptions such as linear contracts or binary effort are often made to avoid this issue. \textcite{edmans2009multiplicative} is an example of an important recent paper using both binary effort and linear contracts.
}

Unfortunately, existing sufficient conditions for the first-order approach to be valid are restrictive. The seminal papers are \textcite{rogerson1985} and \textcite{jewitt1988justifying}, followed by an extensive literature.%
\footnote{Important generalizations include \textcite{sinclair1994first,conlon2009two,jung2015information,chade2020no}, and \textcite{jung2024proxy}. \textcite{chaigneau2022should, chaigneau2024theory} develop sufficient conditions with limited liability.}
\textcite{kadan2017existence} summarize the general view that ``conditions facilitating the first-order approach are typically quite demanding.'' The key issues are elegantly explained by \textcite{chaigneau2022should} and \textcite{conlon2009two}, the latter of whom says ``Unfortunately, the Jewitt conditions are tied to the concavity, not only of the technology, but also of the payment schedule. Thus, there are interesting cases where the Jewitt conditions should fail, since the payment schedule is often not concave. For example, managers often receive stock options, face liquidity constraints.''%
\footnote{\textcite{conlon2009two} continues: ``
I believe that it is natural to expect the first-order approach itself to fail in such cases, since the agent's overall objective function will tend to be nonconcave [...]. The fact that CDFC and CISP imply concavity of the agent's payoff, regardless of the curvature of [the payment], then suggests that the CDFC/CISP conditions are very restrictive, not that the first-order approach is widely applicable.''%
}

Our main result, Theorem~\ref{thm:main}, shows that the first-order approach is broadly valid as
long as the agent's reservation utility is sufficiently high. The first-order approach holds even when
contracts are option-like and the agent's problem has multiple local maxima. Theorem~\ref{thm:main} also implies that the optimal contract exists, is unique, and is characterized by the standard simple formula. There are two main conditions. First, agents must have at least moderate risk aversion. Second, the distribution of output must have $f_{aa}(y|a) \geq 0$ in a sufficiently large region of low scores. The meaning of sufficiently large is made precise in the theorem. Many common distributions have the property that scores are increasing in $y$, and that $f_{aa}(y|a) \geq 0$ in the left tail, so that many common distributions are covered by the theorem (Corollary~\ref{cor:common-utilities}).

\begin{figure}[p]
    \centering
    \includegraphics[width=\textwidth]{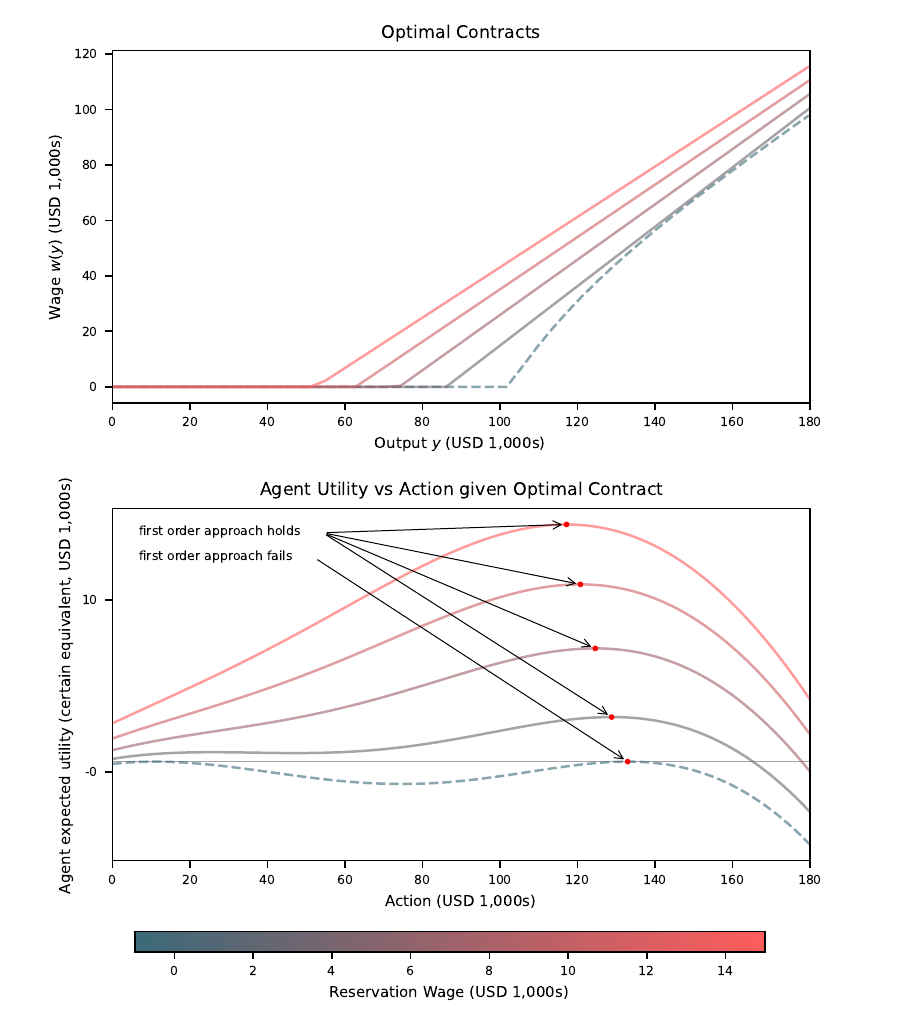}
    \captionsetup{font=footnotesize} % Makes the note footnote-sized
    \caption{Optimal contracts in the principal's problem with Gaussian output and log utility.}
    \label{fig:gaussian-log-pp}
    \caption*{\textit{Note:} Top panel: optimal wage function $w(y)$. Bottom panel: agent's expected utility $U(w^*_{\bar U},a)$ under the displayed contract. For each reservation utility, the displayed contract and recommended action solve the principal's problem. Colors represent reservation utility. Dashed curves indicate that the first-order approach is invalid for the corresponding recommended action and reservation utility. Dots in the bottom panel indicate the recommended action. The thin horizontal line indicates indifference between local maxima. Output is Gaussian with mean $a$ and standard deviation $50$; initial wealth is $50$ (both in thousands of dollars), and the cost function is $c(a) = a^2 / 30000$.}
\end{figure}

The basic idea is illustrated in Figure~\ref{fig:gaussian-log-pp}. When reservation utility is low, the principal offers an option-like contract that pays zero unless output is quite high. This makes the agent's problem non-concave, and the first-order approach fails. However, as reservation utility increases, the kink in the optimal contract moves to the left, and the agent's problem becomes more concave. The first-order approach becomes valid, and the optimal contract is characterized by the standard simple formula.

% Gaussian - log utility figure

Section~\ref{sec:examples} explains our results through simple examples. Section~\ref{sec:model} gives definitions. Section~\ref{sec:main-result} states Theorem~\ref{thm:main}. Section~\ref{sec:main-result-proof} outlines the proof. Section~\ref{sec:calculus} provides closed-form solutions for optimal contracts. Section~\ref{sec:linear-contracts} shows that optimal contracts are piecewise linear option contracts for log utility and output distributions in an exponential family with linear sufficient statistic (this has been independently discovered by \textcite{opp2025moral}). Section~\ref{sec:numerical-methods} provides a state-of-the-art algorithm for finding the optimal contracts. Section~\ref{sec:discussion} discusses the assumptions, results, and connection with the literature. Readers interested in applications may skip to the sufficient conditions in Corollary~\ref{cor:common-utilities} and the applications.

\section{Clarifying Example}
\label{sec:examples}
Consider the following \textbf{Gaussian-log utility example}. A risk-neutral principal hires an agent to perform a task. The agent chooses action $a$ at a cost $c(a)$ proportional to $a^2$. Output $y$, which accrues to the principal, is normally distributed with mean $a$ and standard deviation $\sigma$. The agent has initial wealth $w_0$ and log utility over final wealth, $u(W)=\log W$, where $W=w_0+w(y)$. We consider $a$ on the order of \$100{,}000, a standard deviation $\sigma$ of \$50{,}000, and $w_0$ equal to \$50{,}000. The agent has reservation utility $\bar U$, with certainty-equivalent reservation wage $\bar w$.

The principal designs a compensation contract with wage $w(y) \geq 0$ to maximize profits. The optimal contract balances the goals of inducing effort, reducing the agent's risk, and providing the agent with her reservation utility.

Figure~\ref{fig:gaussian-log-pp} plots optimal contracts for a range of reservation wages $\bar w$, with certainty equivalents from $-\$1,000$ to \$50,000. The top panel plots the optimal wage schedules $w(y)$. The bottom panel plots the agent's expected utility $U(w^*_{\bar U},a)$ under an optimal contract $w^*_{\bar U}$ as a function of action $a$. The intended action is denoted by the red dot in the bottom panel.

The bottom panel shows for what reservation wages the first-order approach is valid. The first-order approach fails whenever there is a binding nonlocal deviation. In these cases, solving the problem with only the local incentive constraint leads to incorrect solutions. Cases in which the first-order approach fails are denoted with dashed lines.

\begin{figure}[p]
    \centering
    \includegraphics[width=\textwidth]{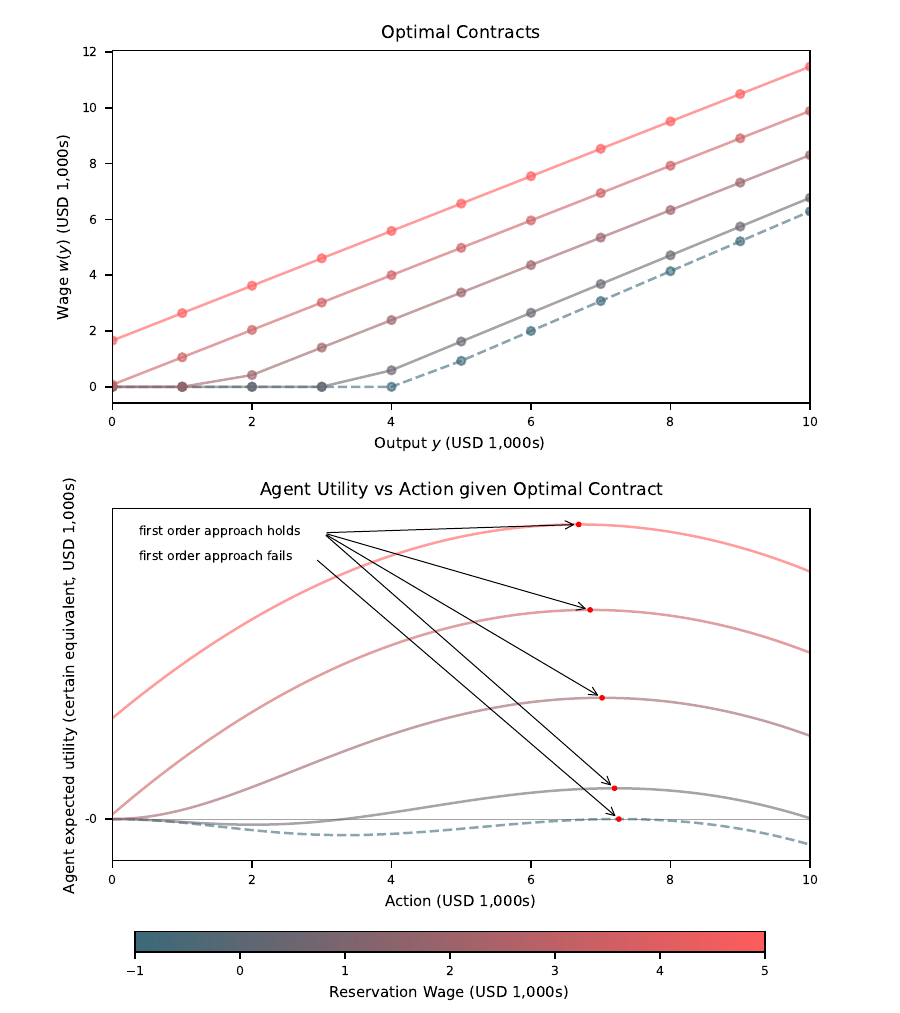}
    \captionsetup{font=footnotesize} % Makes the note footnote-sized
    \caption{Optimal contracts in the principal's problem with Poisson output and log utility.}
    \label{fig:poisson-log-pp}
    \caption*{\textit{Note:} Top panel: optimal wage function $w(y)$. Bottom panel: agent's expected utility $U(w^*_{\bar U},a)$ under the displayed contract. For each reservation utility, the displayed contract and recommended action solve the principal's problem. Colors represent reservation utility. Dashed curves indicate that the first-order approach is invalid for the corresponding recommended action and reservation utility. Dots in the bottom panel indicate the recommended action. The thin horizontal line indicates indifference between local maxima. The number of successes follows a Poisson distribution with mean $a$, and output is one thousand dollars per success. Initial wealth is $50$ (in thousands of dollars), and the cost function is $c(a) = a^2 / 399$.}
\end{figure}

\begin{figure}[p]
    \centering
    \includegraphics[width=\textwidth]{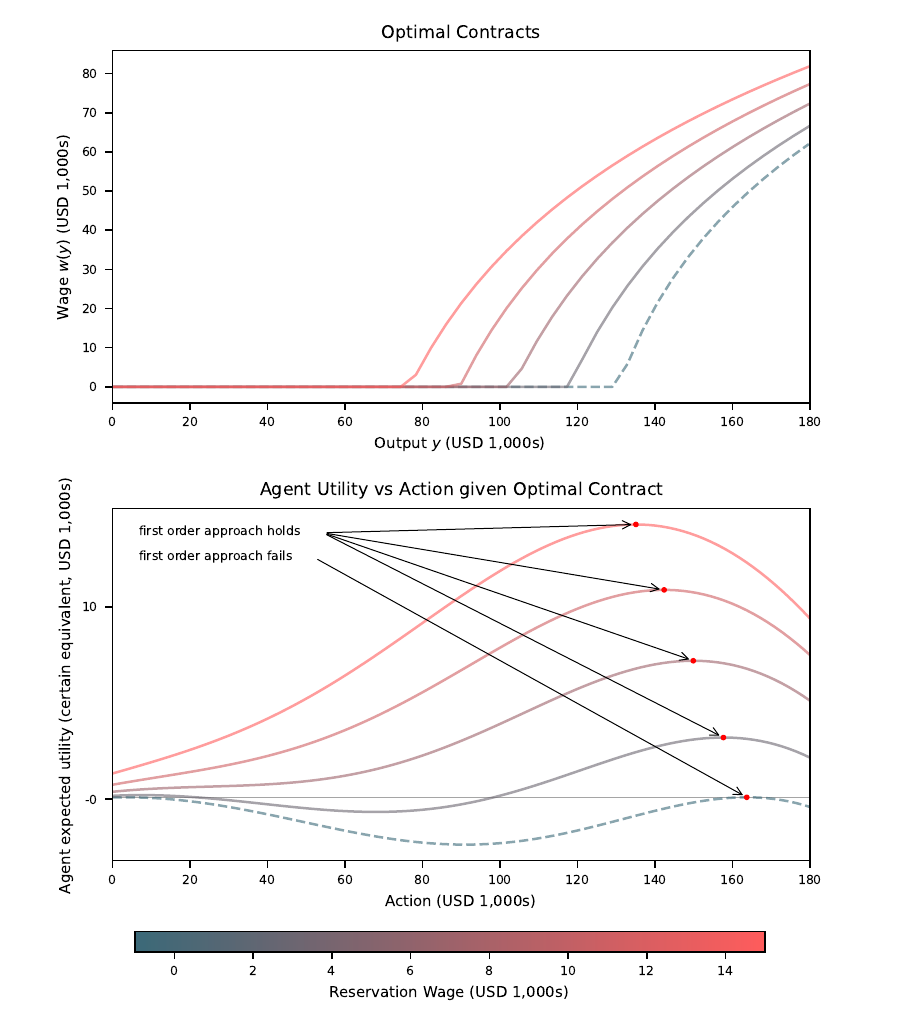}
    \captionsetup{font=footnotesize} % Makes the note footnote-sized
    \caption{Optimal contracts in the principal's problem with Gaussian output and CARA utility.}
    \label{fig:gaussian-cara-pp}
    \caption*{\textit{Note:} Top panel: optimal wage function $w(y)$. Bottom panel: agent's expected utility $U(w^*_{\bar U},a)$ under the displayed contract. For each reservation utility, the displayed contract and recommended action solve the principal's problem. Colors represent reservation utility. Dashed curves indicate that the first-order approach is invalid for the corresponding recommended action and reservation utility. Dots in the bottom panel indicate the recommended action. The thin horizontal line indicates indifference between local maxima. Parameters are as in Figure~\ref{fig:gaussian-log-pp}, but with the CARA coefficient set to yield relative risk aversion of $1$ at the initial wealth.}
\end{figure}

\begin{figure}[p]
    \centering
    \includegraphics[width=\textwidth]{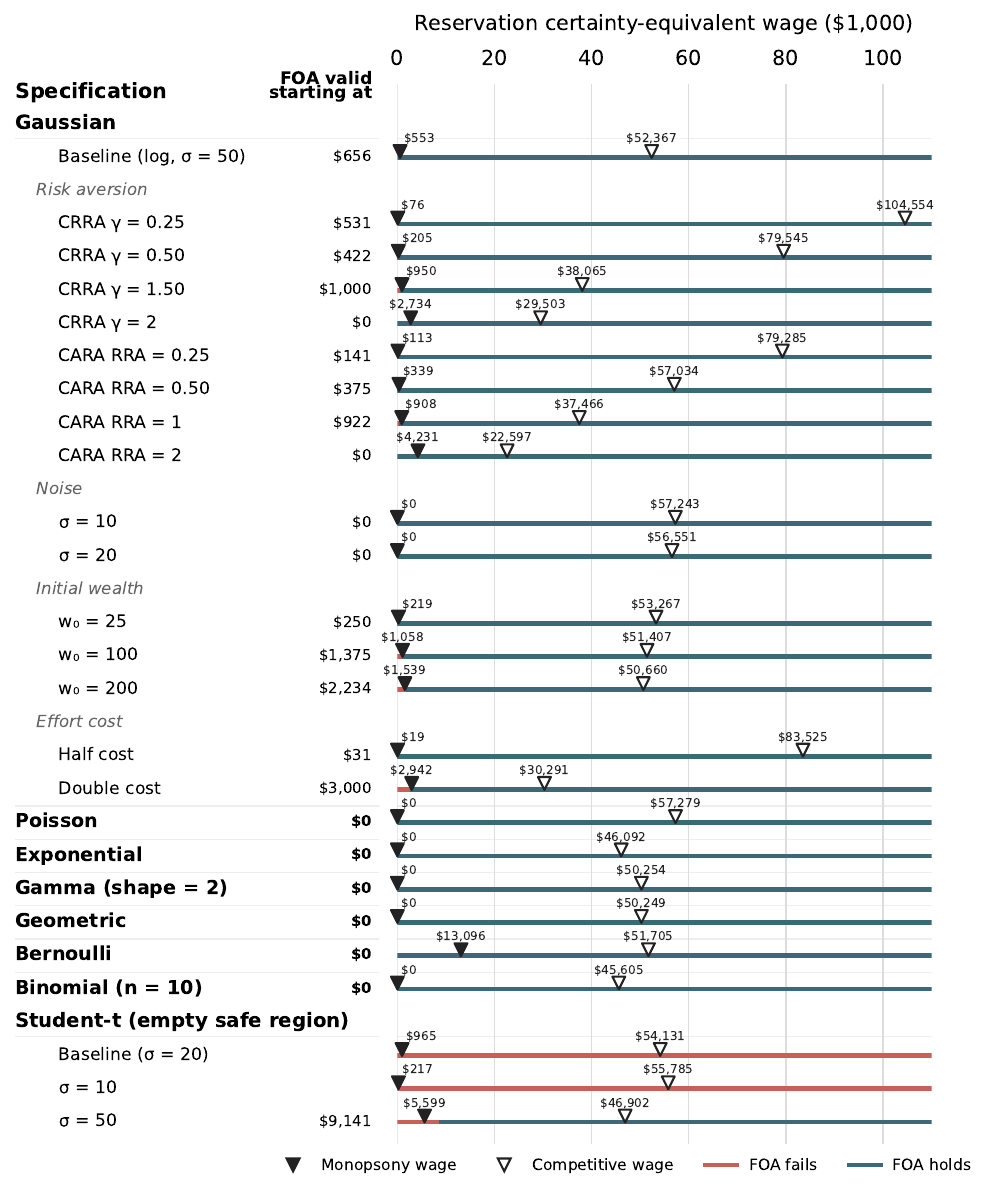}
    \captionsetup{font=footnotesize}
    \caption{Validity of the first-order approach in the principal's problem.}
    \label{fig:foa-principal-summary}
    \caption*{\textit{Note:} Red and teal denote reservation wages at which the first-order approach fails and holds, respectively.  Filled and open triangles mark the monopsony and competitive wages.  Supplementary Appendix~\ref{sec:foa-numerical-experiments} gives the specifications and computational details.}
\end{figure}

The first-order approach fails with a reservation wage of $-\$1,000$, but is valid for all the illustrated positive reservation wages. This is an illustration of our main point, that the first-order approach is broadly valid in most interesting cases. Moreover, this is representative of a more general phenomenon, and not simply this example. Figures~\ref{fig:poisson-log-pp} and \ref{fig:gaussian-cara-pp} consider a different distribution (the discrete Poisson distribution) and a different utility function (CARA). The same pattern holds, as it does in a multitude of other examples. Figure~\ref{fig:foa-principal-summary} summarizes a broader set of numerical experiments. The figure shows that the first-order approach is broadly valid in the experiments. Strikingly, the first-order approach holds for all reservation wages that are at least slightly greater than the monopsony wage. The only exceptions are the Student-$t$ examples, which do not satisfy our requirement that $f_{aa}\geq0$ in a region of low scores. Readers can explore their own examples in the provided \href{https://eduardomazevedo.github.io/azevedo-wolff-2025-site/}{interactive demonstration} and \href{https://colab.research.google.com/github/eduardomazevedo/moralhazard/blob/main/examples/colab_demo.ipynb}{Google Colab notebook}.

% Poisson - log utility figure
\thispagestyle{empty}

% Gaussian - CARA utility figure
\thispagestyle{empty}

% Principal's problem summary figure

Theorem~\ref{thm:main} shows that the first-order approach is valid for sufficiently high reservation utility under commonly satisfied conditions. The examples illustrate the theorem idea in simple terms. For very low reservation utility, the first-order approach often fails. The reason is that the principal offers tough contracts, with zero pay for low output. Optimal contracts have a kink, and only start paying a positive wage relatively close to the intended action. Thus, the agent has a global deviation to basically give up and exert no effort. Thus, the first-order approach fails with very low reservation utility.

However, as \emph{reservation utility increases, the first-order approach becomes valid}. The reason is that the principal is forced to offer contracts with higher utility at the recommended action. This constraint moves the rightmost local maximum up. The rightmost maximum goes up faster than the leftmost local maximum. Thus, the rightmost maximum becomes the unique global maximum. In the illustrative examples, the first-order approach is valid at all displayed positive reservation wages. The proof of Theorem~\ref{thm:main} works with a somewhat different logic. The key observation is that the kink in the relaxed optimal contract moves to the left. This then implies that the agent's problem becomes more concave, and the first-order approach becomes valid.

Theorem~\ref{thm:main} does not contradict existing results in the literature. The low-reservation-utility case confirms the literature's consensus that it is hard to guarantee the validity of the first-order approach \emph{if we insist that it be valid for all reservation utilities}. We see that, even in the Gaussian-log utility case, the first-order approach fails. Thus, any condition guaranteeing the first-order approach for all reservation utilities rules out basic examples and is therefore very restrictive. The examples formally demonstrate the point made by \textcite{chaigneau2022should} and \textcite{conlon2009two}, that requiring the first-order approach to be valid for all reservation utilities excludes many interesting cases.

As a corollary to Theorem~\ref{thm:main}, the moral hazard problem is often tractable. An argument due to \textcite{holmstrom1978incentives} shows that, with the first-order approach, the moral hazard problem has an elegant solution. An optimal contract exists, is unique, and is given by the simple formula
\begin{equation}
    \label{eq:optimal-wage}
    w(y) + w_0=L\bigl(\lambda+\mu s(y)\bigr).
\end{equation}

The wage link function $L$ is determined by the utility function and the limited liability constraint. The score $s(y)$, evaluated at the intended action $a_0$, is determined by the distribution of output. The optimal intended action is $a_0$, and $\lambda$ and $\mu$ are Lagrange multipliers. We give closed-form solutions for these functions for a wide range of examples, and use these formulas to develop an algorithm to compute optimal contracts.

Another corollary is that optimal contracts are piecewise linear in a variety of examples. With log utility, the wage link function is piecewise linear. With output distributions in an exponential family with linear sufficient statistic, the score is linear in $y$. Thus, with both of these conditions, the optimal contract is piecewise linear. Figures~\ref{fig:gaussian-log-pp} and \ref{fig:poisson-log-pp} illustrate this in the Gaussian and Poisson cases. Note that the optimal contract is only piecewise linear when the first-order approach is valid. This is clear in Figure~\ref{fig:gaussian-log-pp}, where the optimal contract is noticeably nonlinear for a reservation wage of $-\$1,000$, but otherwise piecewise linear. The CARA case in Figure~\ref{fig:gaussian-cara-pp} illustrates that, without log utility, the optimal contract is not piecewise linear.

\section{Model}
\label{sec:model}
\subsection{Model}

A risk-neutral \textbf{principal} hires an \textbf{agent} to take an action
$a$ in a compact interval $\mathcal A\subseteq\mathbb R_+$.  \textbf{Output}
$y$ lies in a
common support $\mathcal Y$ and has a strictly positive density or probability
mass function $f(y\mid a)$ with respect to a common dominating measure $\nu$.
Thus the model includes continuous and discrete output.

The agent has initial wealth $w_0$ and utility $u(W)$ over final wealth
$W\in[w_0,\infty)$.  A \textbf{contract} is a measurable payment function
$w:\mathcal Y\to\mathbb R_+$; the restriction $w(y)\geq0$ is limited liability.
Under contract $w$ and action $a$, the agent's expected utility and the
principal's expected compensation cost are
\begin{align*}
    U(w,a)
    &:=\int u\bigl(w_0+w(y)\bigr)f(y\mid a)\,d\nu(y)-c(a),\\
    C(w,a)
    &:=\int w(y)f(y\mid a)\,d\nu(y),
\end{align*}
where $c(a)$ is the agent's cost of action.  Let $\mathcal W$ denote the set of
feasible contracts, specified below.

Throughout the paper, we fix an interior
\textbf{intended action} $a_0\in\mathcal A$.
Given reservation utility
$\bar U$, the \textbf{cost-minimization problem}
chooses $w\in\mathcal W$ to
\begin{align}
    \text{minimize}\quad & C(w,a_0) \nonumber\\
    \text{subject to}\quad
    &U(w,a_0)\geq\bar U, \tag{IR}\label{IR}\\
    &U(w,a_0)\geq U(w,a)
      \quad\text{for every }a\in\mathcal A. \tag{GIC}\label{GIC}
\end{align}
Global incentive compatibility requires that the intended action be optimal for
the agent.

The \textbf{relaxed cost-minimization problem} replaces global incentive
compatibility with the weaker local incentive compatibility condition
\begin{equation}
    U_a(w,a_0)=0. \tag{LIC}\label{LIC_relaxed}
\end{equation}

We say that the \textbf{first-order approach is valid} for $(a_0,\bar U)$ if
every solution of the relaxed problem also solves the original
cost-minimization problem.

Throughout the paper, we always fix the intended action and focus on cost minimization unless otherwise specified. For some examples and discussion we consider the \textbf{principal's problem} to choose both the intended action and contract to maximize
\[
    \Revenue(a_0)-C(w,a_0),
\]
for a given \textbf{revenue} function $\Revenue(a)$, subject to the same constraints as the cost-minimization problem. The \textbf{relaxed principal's problem} replaces the global incentive compatibility constraint with the local incentive compatibility constraint.

\subsection{Assumptions and notation}
\textbf{Maintained assumptions.}
We maintain the following assumptions throughout the paper.  For
$\nu$-almost every $y$, the map $a\mapsto f(y\mid a)$ is twice continuously
differentiable.  Utility is three times continuously differentiable, with
$u'(W)>0$, $u''(W)<0$, and $\lim_{W\to\infty}u'(W)=0$.  The cost function $c$
is twice continuously differentiable and strictly convex, with $c'(a_0)>0$.
Contracts in $\mathcal W$ have finite expected compensation cost and expected
utility at every action and permit differentiation under the integral sign
through second order.  Thus, for $j=1,2$,
\begin{equation}
    \frac{d^j}{da^j}\int
    u\bigl(w_0+w(y)\bigr)f(y\mid a)\,d\nu(y)
    =
    \int u\bigl(w_0+w(y)\bigr)
    \partial_a^j f(y\mid a)\,d\nu(y).
    \label{eq:contract-differentiation}
\end{equation}
Every canonical contract in Definition~\ref{def:canonical-contract} belongs to $\mathcal W$, and its
expected utility, local incentives, and expected compensation cost vary
continuously with the canonical-contract parameters.

Finally, there is enough utility variation to provide the intended local
incentives:
\begin{equation}
    \left[\lim_{W\to\infty}u(W)-u(w_0)\right]
    \int_{\{f_a(y\mid a_0)>0\}}f_a(y\mid a_0)\,d\nu(y)
    >c'(a_0).
    \label{eq:local-implementability}
\end{equation}

\textbf{Notation.}
We use the following notation.  The \textbf{score at the intended action} is
\[
    s(y):=\frac{f_a(y\mid a_0)}{f(y\mid a_0)}.
\]

Let $U_R$ be the supremum reservation utility for which the relaxed
cost-minimization problem is feasible, and let
$\underline c'' := \inf_{a\in\mathcal A}c''(a)\geq0$.

The main theorem depends on regions where the output distribution has
$f_{aa}\geq0$.  For $q\leq0$, call $\{y:s(y)\leq q\}$ a \textbf{safe
low-score region} if $f_{aa}(y\mid a)\geq0$ for every $a\in\mathcal A$ and
$\nu$-almost every $y$ in that region.  Let $\mathcal Q_{\mathrm{safe}}$ be the
set of cutoffs that generate safe low-score regions, and let
$\bar q_{\mathrm{safe}}:=\sup\mathcal Q_{\mathrm{safe}}$.  The \textbf{safe
incentive capacity} is
\begin{equation}
    \mathrm{Cap}_{\mathrm{safe}}
    :=-
    \int_{\{s(y)<\bar q_{\mathrm{safe}}\}}
    s(y)f(y\mid a_0)\,d\nu(y).
    \label{eq:safe-capacity}
\end{equation}
It measures how much incentive power is available from outcomes in the safe
low-score region.  The positive curvature outside that region is bounded by the
\textbf{safe curvature},
\begin{equation}
    \mathrm{Curv}_{\mathrm{safe}}
    :=\inf_{q\in\mathcal Q_{\mathrm{safe}}}
      \sup_{a\in\mathcal A}
      \int_{\{s(y)>q\}}
      \bigl[s(y)f_{aa}(y\mid a)\bigr]_+\,d\nu(y).
    \label{eq:safe-curvature}
\end{equation}

The \textbf{safe width} is
\begin{equation}
    \mathrm{Width}_{\mathrm{safe}}
    :=\sup\left(\{0\}\cup
    \left\{q-\operatorname*{ess\,inf}_{y\in\mathcal Y}s(y):
    q\in\mathcal Q_{\mathrm{safe}}\right\}\right).
    \label{eq:safe-width}
\end{equation}
In the curvature--width ratio below, the ratio is zero when curvature is finite
and width is infinite, and infinite when curvature is infinite or width is zero.

\begin{remark}[Safe tails in common distributions]
\label{rem:safe-tails-common-distributions}
For Gaussian output (fixed positive variance); Poisson and exponential output
($0\notin\mathcal A$); gamma output (fixed positive shape and $0\notin\mathcal A$);
geometric output ($\mathcal A\subset(1,\infty)$); and Bernoulli and binomial output
($\mathcal A\subset(0,1)$, with a fixed number of trials),
\[
    \mathcal Q_{\mathrm{safe}}\neq\varnothing,
    \qquad \mathrm{Cap}_{\mathrm{safe}}>0,
    \qquad \mathrm{Curv}_{\mathrm{safe}}<\infty,
    \qquad \mathrm{Width}_{\mathrm{safe}}>0.
\]
Gaussian has infinite safe width.  The same conclusions, with
infinite safe width, hold for a regular exponential family with an affine
sufficient statistic that is essentially unbounded below, provided its natural
parameter $\eta(a)$ is twice continuously differentiable,
$\inf_{a\in\mathcal A}\eta'(a)>0$, and $\eta(\mathcal A)$ lies in the interior
of the natural-parameter space.  Student's $t$, by contrast, need not have
positive safe width.
\end{remark}

The following utility condition is used for part of the main theorem.

\begin{condition}[Sufficient risk aversion]
\label{condition:sufficient-risk-aversion}
Let $A(W):=-u''(W)/u'(W)$ denote absolute risk aversion and
$P(W):=-u'''(W)/u''(W)$ denote absolute prudence.  For every $W>w_0$,
\begin{alignat}{2}
    &\frac{A(W)}{P(W)}
    &&\geq \frac{1}{3},
    \tag{U1}\label{condition:U1}\\
    \lim_{W\to\infty}\quad&\frac{A(W)}{u'(W)}
    &&=\infty.
    \tag{U2}\label{condition:U2}
\end{alignat}
\end{condition}

\begin{remark}[Sufficient risk aversion condition in common utility functions]
The sufficient risk aversion condition is satisfied by CARA utility and by CRRA utility with coefficient of relative risk aversion strictly greater than $1$.
\end{remark}

The preceding remarks are verified in Supplementary Appendix~\ref{sec:common-regularity-verification}.

\section{Main Result}
\label{sec:main-result}
The main result gives conditions under which the first-order approach is valid
at every sufficiently high feasible reservation utility.

\begin{theorem}[The first-order approach is valid at sufficiently high reservation utility]
\label{thm:main}
Suppose either:
\begin{enumerate}
    \renewcommand{\labelenumi}{(\roman{enumi})}
    \item utility satisfies Condition~\ref{condition:sufficient-risk-aversion}, $\underline c''>0$,
    $\mathrm{Curv}_{\mathrm{safe}}<\infty$, and
    \[
        \left[\lim_{W\to\infty}u(W)-u(w_0)\right]
        \mathrm{Cap}_{\mathrm{safe}}>c'(a_0);
    \]
    \item $u(W)=\log W$ and
    \begin{equation}
        \frac{\mathrm{Curv}_{\mathrm{safe}}}
        {\mathrm{Width}_{\mathrm{safe}}}
        <\underline c''.
        \label{eq:log-curvature-width-condition}
    \end{equation}
\end{enumerate}
Then there exists $U^*<U_R$ such that, for every
\[
    \bar U\in[U^*,U_R),
\]
the first-order approach is valid. The relaxed cost-minimization problem has an almost everywhere unique solution characterized in Proposition~\ref{prop:relaxed-optimal-contract}.
\end{theorem}

The theorem immediately implies the following sufficient conditions for common utility specifications. Part (iii) is based on the extension in Supplementary Appendix~\ref{sec:asymptotic-affinity}. See Supplementary Appendix~\ref{sec:proof-common-utilities} for the proof.

\begin{corollary}[Common Utility Specifications]
\label{cor:common-utilities}
Assume $\underline c''>0$.
\begin{enumerate}
    \renewcommand{\labelenumi}{(\roman{enumi})}
    \item Suppose $\mathrm{Cap}_{\mathrm{safe}}>0$ and
    $\mathrm{Curv}_{\mathrm{safe}}<\infty$.  For CARA utility or CRRA utility
    with coefficient of relative risk aversion $\gamma>1$, the first-order
    approach is valid at all sufficiently high feasible reservation utilities
    whenever initial wealth is sufficiently low.  In particular, this applies
    to Gaussian, Poisson, exponential, gamma, geometric, Bernoulli, and binomial
    output under the parameter restrictions in Remark~\ref{rem:safe-tails-common-distributions}.

    \item For log utility with $w_0>0$, if the score is essentially unbounded
    below and $\mathrm{Curv}_{\mathrm{safe}}<\infty$, the first-order approach
    is valid for sufficiently high reservation utility.  In particular, this
    applies to Gaussian output and to the regular exponential
    families described in Remark~\ref{rem:safe-tails-common-distributions}.

    \item For CRRA utility with $w_0>0$ and coefficient of relative risk
    aversion $\gamma\in(0,1)$, the first-order approach is
    valid at all sufficiently high feasible reservation utilities for Gaussian,
    Poisson, exponential, gamma, geometric, Bernoulli, and binomial output under
    the parameter restrictions in Remark~\ref{rem:safe-tails-common-distributions}.
\end{enumerate}
\end{corollary}

\section{Proof of the Main Result}
\label{sec:main-result-proof}
The proof has two steps.  First, Proposition~\ref{prop:relaxed-optimal-contract} characterizes the solution of the relaxed
problem.  Second, we show that, under the conditions of Theorem~\ref{thm:main},
the agent's objective under that contract is concave in action at every
sufficiently high feasible reservation utility.  Local incentive compatibility
then implies global incentive compatibility.

\subsection{Solution to the Relaxed Problem}

We first show that the relaxed cost-minimization problem has an almost
everywhere unique solution with a simple representation in terms of Lagrange
multipliers.  This is a minor extension of standard results on the relaxed
problem going back to \textcite{holmstrom1978incentives} and
\textcite{jewitt2008moral}.

Let $\lambda\geq0$ be the multiplier on individual rationality and let $\mu$ be
the multiplier on local incentive compatibility.  The Lagrangian is
\begin{equation}
    \mathcal L(w,\lambda,\mu)
    :=C(w,a_0)+\lambda\bigl[\bar U-U(w,a_0)\bigr]
      -\mu U_a(w,a_0).
    \label{eq:lagrangian}
\end{equation}
Heuristically differentiating pointwise with respect to $w(y)$
gives, whenever limited liability does not
bind,
\begin{equation}
    \frac{1}{u'\bigl(w_0+w(y)\bigr)}=\lambda+\mu s(y).
    \tag{FOC}\label{eq:foc}
\end{equation}

The first-order condition pins down the optimal contract $w^*_{\bar U}(y)$.
The solution is best described with the following notation.  Let
$m_0:=1/u'(w_0)$ be the marginal cost of utility at zero payment.  On the
nonbinding region, the solution to the first-order condition is calculated by taking $\lambda + \mu s(y)$ and applying the \textbf{base wage link function}
\[
    L_0(z):=(u')^{-1}(1/z).
\]
Taking limited liability into account, we define the \textbf{wage link function}
\[
    L(z):=L_0(z\vee m_0).
\]

The optimal contract can be cleanly expressed as follows.

\begin{definition}[Canonical contract]
\label{def:canonical-contract}
Given multipliers $\lambda,\mu\in\mathbb R$, the associated \textbf{canonical contract} is
\begin{equation}
    w_{\lambda,\mu}(y):=L\bigl(\lambda+\mu s(y)\bigr)-w_0.
    \label{eq:canonical-contract}
\end{equation}
\end{definition}

The following standard proposition characterizes the relaxed solution.

\begin{proposition}[Solution of the Relaxed Cost-Minimization Problem]
\label{prop:relaxed-optimal-contract}
\textnormal{\parencite{holmstrom1978incentives,jewitt2008moral}}
There exists $U_L<U_R$ such that, for every $\bar U<U_R$:
\begin{enumerate}
    \renewcommand{\labelenumi}{(\roman{enumi})}
    \item The relaxed problem has an almost everywhere unique solution
    $w^*_{\bar U}$, which is almost everywhere a canonical contract for some multipliers $\lambda,\mu$:
    \begin{equation}
        w^*_{\bar U}(y)=w_{\lambda,\mu}(y).
        \label{eq:relaxed-optimal-contract}
    \end{equation}

    \item If $\bar U\leq U_L$, the relaxed-optimal contract
    is constant in $\bar U$.  If $U_L<\bar U<U_R$, individual
    rationality binds, every IR multiplier $\lambda$ associated with a higher reservation
    utility exceeds every IR multiplier associated with a lower one, and the
    smallest associated multiplier converges to infinity as $\bar U\uparrow
    U_R$.

    \item The
    \textbf{optimal expected wage cost} $C^*(\bar U)$ is constant below $U_L$ and
    strictly increasing and strictly convex on $(U_L,U_R)$.
\end{enumerate}
\end{proposition}

\begin{figure}[ht]
    \centering
    \includegraphics{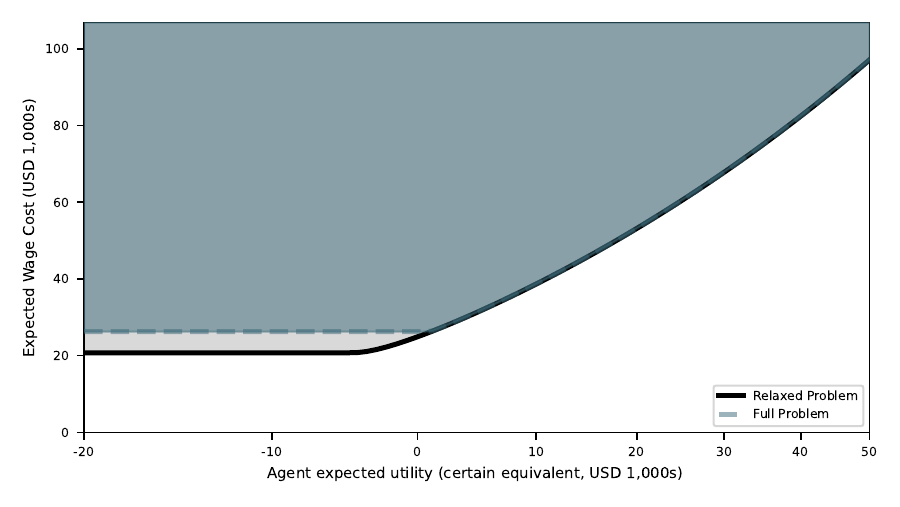}
    \caption{Pareto frontiers in the Gaussian-log cost-minimization problem.}
    \label{fig:pf}
    \captionsetup{font=footnotesize} % Makes the note footnote-sized
    \caption*{\textit{Note:} This figure displays the frontiers of the cost-minimization problem and its relaxation for the fixed intended action $a_0=\$100{,}000$. The first-order approach is valid at reservation utilities where the two frontiers coincide. Other parameters are as in Figure~\ref{fig:gaussian-log-pp}.}
\end{figure}

The Gaussian-log utility example illustrates Proposition~\ref{prop:relaxed-optimal-contract}.  Here
$s(y)=(y-a_0)/\sigma^2$ and $L(z)=z\vee w_0$, so the relaxed-optimal wage is
\[
    w^*_{\bar U}(y)
    =\left[\lambda+\mu\frac{y-a_0}{\sigma^2}-w_0\right]_+.
\]
This is the piecewise-linear solution seen in Figure~\ref{fig:gaussian-log-pp}.  Figure~\ref{fig:pf} illustrates the associated
cost--utility frontier.

% Pareto frontier figure

\subsection{Validity of the First-Order Approach with High Reservation Utility}

We prove the
stronger result that the agent's objective is concave in $a$.

\begin{proposition}[Concavity at High Reservation Utility]
\label{prop:concave}
Under either set of conditions in Theorem~\ref{thm:main}, there exists
$\lambda_0<\infty$ such that every canonical contract $w_{\lambda,\mu}$ that
satisfies local incentive compatibility and has $\lambda\geq\lambda_0$ obeys
\[
    U_{aa}(w_{\lambda,\mu},a)\leq0
    \qquad\text{for every }a\in\mathcal A.
\]
\end{proposition}

Theorem~\ref{thm:main} follows directly from Proposition~\ref{prop:concave}:

\begin{proof}[Proof of Theorem~\ref{thm:main}]
Take $\lambda_0$ as in Proposition~\ref{prop:concave}.  Proposition~\ref{prop:relaxed-optimal-contract} implies that the smallest multiplier
supporting the relaxed-optimal contract converges to infinity as
$\bar U\uparrow U_R$.  Hence there is $U^*<U_R$ such that every multiplier
supporting the relaxed solution has $\lambda\geq\lambda_0$ whenever
$\bar U\in[U^*,U_R)$.  The agent's objective is then concave by Proposition~\ref{prop:concave}.  Local incentive compatibility implies global incentive compatibility, so the unique
relaxed solution also uniquely solves the original cost-minimization problem.
\end{proof}

The proof of Proposition~\ref{prop:concave} is essentially a three-line
argument that we outline here.  The first step is to show that, as $\lambda$ grows, the
limited-liability kink in a locally incentive-compatible canonical contract
moves into a low-score region (Lemma~\ref{lem:kink-placement}).  To see this,
consider the local incentive-compatibility condition
\[
    c'(a_0)
    =\int u\bigl(L(\lambda+\mu s(y))\bigr)s(y)
      f(y\mid a_0)\,d\nu(y).
\]
Because the score has mean zero, we can add and subtract $u(L(\lambda))$ to
obtain
\[
    c'(a_0)
    =\int\bigl[u\bigl(L(\lambda+\mu s(y))\bigr)
                       -u\bigl(L(\lambda)\bigr)\bigr]s(y)
      f(y\mid a_0)\,d\nu(y).
\]
Suppose that a cutoff $q_0<0$ lies weakly to the left of the limited-liability
kink, so wages are
zero whenever $s(y)\leq q_0$.  The local incentive multiplier satisfies
$\mu>0$, and the wage link function is nondecreasing, so the utility difference and
the score have the same weak sign.  The centered integrand is therefore
nonnegative at every score, and
\begin{align*}
    c'(a_0)
    &\geq \int_{\{s(y)\leq q_0\}}
       \bigl[u(w_0)-u\bigl(L(\lambda)\bigr)\bigr]s(y)
       f(y\mid a_0)\,d\nu(y)\\
    &=\bigl[u\bigl(L(\lambda)\bigr)-u(w_0)\bigr]
      \left[-\int_{\{s(y)\leq q_0\}}s(y)f(y\mid a_0)\,d\nu(y)\right].
\end{align*}
As $\lambda$ grows, the first factor on the right grows.  The capacity
assumptions of the theorem then guarantee that we can find a safe cutoff $q_0$
for which the right-hand side exceeds $c'(a_0)$ for all sufficiently large
$\lambda$.  Thus, the limited-liability kink must eventually lie to the left
of $q_0$.

The second step is to show that only curvature above a safe cutoff $q_1 > q_0$ can
work against concavity.  Since $f_{aa}$ has integral zero, we can center
the agent's objective to obtain
\[
    U_{aa}(w_{\lambda,\mu},a)
    =\int\!\left[u\bigl(L(\lambda+\mu s(y))\bigr)
                  -u\bigl(L(\lambda)\bigr)\right]
      f_{aa}(y\mid a)\,d\nu(y)-c''(a).
\]
On $\{s(y)\leq q_1\}$, where $q_1\leq0$ is safe, the utility difference is
nonpositive and $f_{aa}(y\mid a)\geq0$.  This region therefore contributes
weakly negatively and can be dropped when deriving an upper bound.  It remains
only to bound the integral over $\{s(y)>q_1\}$.

The third step bounds this remaining curvature.  The fact that the kink is left of $q_0$ prevents the
incentive multiplier $\mu$ from growing too quickly relative to $\lambda$.
The utility conditions then tame the sensitivity of utility to the score on
the region above $q_1$, while the safe-curvature bound limits the positive
distributional curvature in that region.  Using these facts, Proposition~\ref{prop:finite-lambda-bound} shows that the remaining curvature is small, and is eventually dominated by $\underline c''$.

\begin{figure}[p]
    \centering
    \includegraphics[width=\textwidth]{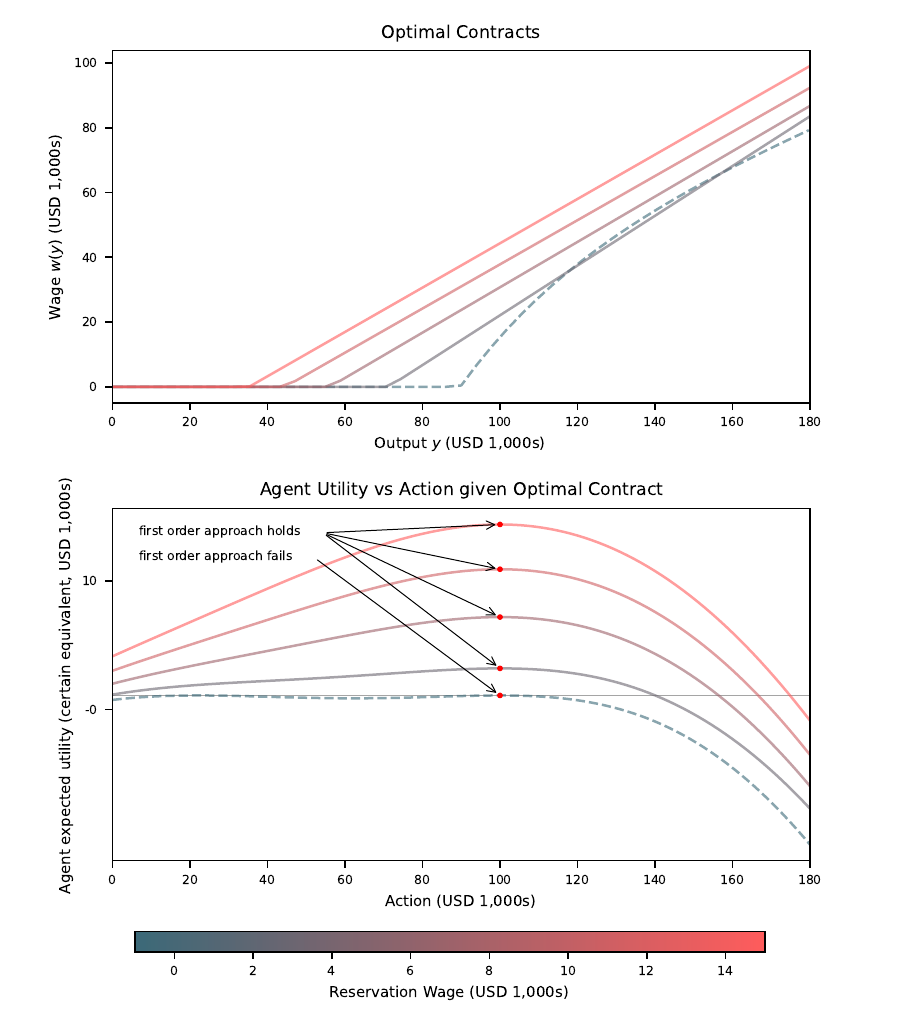}
    \captionsetup{font=footnotesize}
    \caption{Optimal contracts with Gaussian output and log utility in the
    cost-minimization problem with intended action $a_0=\$100{,}000$.}
    \label{fig:gaussian-log-cm}
    \caption*{\textit{Note:} Top panel: optimal wage function $w(y)$. Bottom
    panel: agent's expected utility $U(w^*_{\bar U},a)$ under the displayed
    contract. Colors represent reservation utility. Dashed curves indicate that
    the first-order approach is invalid for the fixed intended action at the
    corresponding reservation utility. Dots in the bottom panel indicate the
    intended action. The thin horizontal line indicates indifference between
    local maxima. Other parameters are as in Figure~\ref{fig:gaussian-log-pp}.}
\end{figure}

Figure~\ref{fig:gaussian-log-cm} illustrates the argument in the log-Gaussian example.  At low reservation
utility, the limited-liability kink creates multiple local maxima.  As
reservation utility rises, the kink moves into the safe left tail as predicted
by Lemma~\ref{lem:kink-placement} (top panel of the figure).  As the kink moves far to the left, the agent's objective becomes concave
as predicted by Proposition~\ref{prop:finite-lambda-bound} (bottom panel of the figure).  Appendix~\ref{subsec:appendix-safe-tail} gives the proof.  While care is needed to deal with the continuous and discrete cases, and the two branches of Theorem~\ref{thm:main}, the gist is the three-line argument above.

% Gaussian - log utility cost-minimization problem figure

\section{Applications}
\label{sec:applications}
\subsection{Closed-Form Solutions}
\label{sec:calculus}
Theorem~\ref{thm:main} implies that optimal contracts have simple functional forms in parametric settings. Contracts depend on the utility function and limited liability constraint through the wage link function $L$, and on the distribution of output through the score $s(y)$. Table~\ref{tab:utility_functions} gives the wage link function for common utility functions, while Table~\ref{tab:dists} gives action scores for common output distributions. Derivations are in the supplementary material.

\begin{table}[ht]
    \centering
    \caption{Utility Functions and Wage Link Functions}
    \label{tab:utility_functions}
    \renewcommand{\arraystretch}{1.5}
    \setlength{\tabcolsep}{10pt}
    \begin{tabular*}{\textwidth}{@{\extracolsep{\fill}}lcc}
        \toprule
        & \multicolumn{1}{c}{Utility Function}
        & \multicolumn{1}{c}{Wage Link Function} \\
        & \(u(W)\) & \(L(z)\) \\
        \midrule
        Log
        & \(\log W\)
        & \(z\vee w_0\) \\
        CRRA
        & \(\dfrac{W^{1-\gamma}}{1-\gamma}\)
        & \(\bigl(z\vee w_0^\gamma\bigr)^{1/\gamma}\) \\
        CARA
        & \(-\dfrac{\exp(-\alpha W)}{\alpha}\)
        & \(\dfrac{\log\bigl(z\vee\exp(\alpha w_0)\bigr)}{\alpha}\) \\
        \bottomrule
    \end{tabular*}
    \captionsetup{font=footnotesize}
    \caption*{\textit{Note:} Utility $u$ is defined over total wealth $W$.  The
    wage link function $L$ gives canonical total wealth; subtracting initial wealth gives
    the associated wage,
    $w_{\lambda,\mu}(y)=L(\lambda+\mu s(y))-w_0$.  For CRRA,
    $\gamma>0$ and $\gamma\neq1$; for CARA, $\alpha>0$.}
\end{table}

\begin{table}[ht]
    \centering
    \caption{Output Distributions and Scores}
    \label{tab:dists}
    \renewcommand{\arraystretch}{1.5} % Reduce vertical spacing
    \small % Reduce font size further
    \setlength{\tabcolsep}{6pt} % Reduce column spacing for compactness
    \resizebox{\textwidth}{!}{ % Ensures table fits within page width
    \begin{tabular}{l p{5.5cm} p{3.8cm} p{3.8cm}} % Keep all columns aligned and tight
        \toprule
        \textbf{Distribution}
        & \multicolumn{1}{l}{\textbf{\(f(y\mid a)\)}}
        & \multicolumn{1}{l}{\textbf{Score} \(\partial_a\log f(y\mid a)\)}
        & \textbf{Mean} \\
        \midrule
        Gaussian
        & \( \mathcal{N}(a, \sigma^2) \)
        & \( \frac{y - a}{\sigma^2} \) 
        & \( a \) 
        \\ 

        Lognormal
        & \( \frac{1}{y \sqrt{2\pi\sigma^2}} \exp \!\Bigl( -\frac{(\log(y)-a)^2}{2\sigma^2} \Bigr) \) 
        & \( \frac{\log(y) - a}{\sigma^2} \) 
        & \( \exp\!\Bigl(a + \frac{\sigma^2}{2}\Bigr) \) 
        \\ 

        Poisson
        & \( \frac{a^y e^{-a}}{y!} \) 
        & \( \frac{y - a}{a} \) 
        & \( a \) 
        \\ 

        Exponential 
        & \( \frac{1}{a} e^{-\tfrac{y}{a}} \) 
        & \( \frac{y-a}{a^2} \) 
        & \( a \) 
        \\ 

        Bernoulli
        & \( a^y (1 - a)^{1 - y} \), \( y \in \{0, 1\} \) 
        & \( \frac{y-a}{a-a^2} \)
        & \( a \) 
        \\  

        Geometric
        & \( \Bigl(1 - \frac{1}{a}\Bigr)^{y - 1} \Bigl(\frac{1}{a}\Bigr) \), \( y \in \{1,2,\dots\} \) 
        & \( \frac{y - a}{a^2 - a} \) 
        & \( a \) 
        \\ 

        Binomial
        & \( \binom{n}{y} a^y (1 - a)^{n - y} \), \( y \in \{0,\dots,n\} \) 
        & \( \frac{y - na}{a - a^2} \) 
        & \( n a \) 
        \\ 

        Gamma
        & \( f(y \mid n, a) = \frac{y^{n - 1} e^{-\tfrac{y}{a}}}{\Gamma(n)\, a^n} \) 
        & \( \frac{y - n a}{a^2} \) 
        & \( n a \)
        \\ 

        Student's \(t\)
        & \( 
        \frac{\Gamma\!\bigl(\tfrac{d + 1}{2}\bigr)}{\Gamma\!\bigl(\tfrac{d}{2}\bigr)\,\sqrt{\pi d}\,\sigma}
        \left(1 + \frac{1}{d} \,\frac{(y - a)^2}{\sigma^2}\right)^{-\tfrac{d + 1}{2}}
        \) 
        & \( \frac{(d+1)(y-a)}{d\,\sigma^2 + (y-a)^2} \)
        & \( a \) 
        \\ 

        Exponential Family
        & \( h(y)\,\exp\Bigl(\eta(a)\,T(y) - \psi(a)\Bigr) \)
        & \( T(y)\,\frac{d\eta(a)}{da} \;-\; \frac{d\psi(a)}{da} \)
        & \textit{(Not specified)} 
        \\ 

        \( y = a + X, X \sim h \)
        & \( h(y - a) \)
        & \( -\frac{h'(y - a)}{h(y - a)} \)
        & \( a + \mathbb{E}[X] \) 
        \\ 

        \( y = a X, X \sim h \)
        & \( \bigl|\tfrac{1}{a}\bigr|\;h\!\Bigl(\tfrac{y}{a}\Bigr) \)
        & \( -\frac{1}{a} - \frac{y}{a^2} \frac{h'(\frac{y}{a})}{h(\frac{y}{a})} \)
        & \( a \,\mathbb{E}[X] \)
        \\ 
        \bottomrule
    \end{tabular}
    } % End resizebox
    \captionsetup{font=footnotesize} % Match figure's note formatting
    \caption*{\textit{Note:} The table reports the probability density function (or probability mass function for discrete distributions), action score, and mean as functions of the agent's action $a$.  For Student's $t$, $d$ denotes degrees of freedom.}
\end{table}

\subsection{Linear Contracts}
\label{sec:linear-contracts}
The formulae imply piecewise linear contracts in many examples.  With log
utility, $L(z)=z\vee w_0$. So, if the score is affine in output, the canonical wage is piecewise linear. This
includes an exponential family with a linear sufficient statistic.  That is,
suppose $f(y\mid a)$ is of the form
\begin{equation}
    f(y\mid a)
    =h(y)\exp\bigl(\eta(a)y-\psi(a)\bigr).
    \label{eq:linear-exponential-family}
\end{equation}
Its score at the intended action is
\[
    s(y)=\eta'(a_0)y-\psi'(a_0),
\]
which is affine in $y$.

We note this as follows:

\begin{remark}[Piecewise Linear Option Contracts]
    \label{rem:linear-contracts}
    Assume that utility is log ($u(W)=\log W$ with $w_0>0$) and that the distribution of outcomes is in an exponential family with a linear sufficient statistic as in equation~\eqref{eq:linear-exponential-family}. Then the canonical contract wage function is a piecewise linear option contract. This includes the Gaussian, exponential, Poisson, geometric, binomial, and gamma distributions.
\end{remark}

Our piecewise linear contracts are option contracts. With increasing score, the contracts pay zero for low outcomes and increase linearly past a ``strike price.'' In some examples with a lower bound on the support, the optimal contract is globally linear (e.g., some of the Poisson examples in Figure~\ref{fig:poisson-log-pp}). Finally, note that optimal contracts are only piecewise linear when the first-order approach is valid. This is illustrated in the Gaussian-log utility example in Figure~\ref{fig:gaussian-log-pp}, where the contract with negative reservation wage is clearly nonlinear.

\subsection{Numerical Methods}
\label{sec:numerical-methods}
Theorem~\ref{thm:main} and the closed-form solutions yield efficient algorithms for solving the cost-minimization problem.

Consider a relaxed cost-minimization problem in which we also include a finite number of global incentive compatibility constraints at a vector of actions $\boldsymbol{\hat a} = (\hat a_1, \ldots, \hat a_n)$. The Lagrangian from equation~\eqref{eq:lagrangian} becomes
\[
    \begin{aligned}
    \mathcal{L}(w,\lambda,\mu, \boldsymbol{\hat \mu}, \boldsymbol{\hat a})
    &:=
    C(w, a_0)
    +\lambda\left(\bar{U}-U(w, a_0)\right)
    +\mu(-U_{a}(w, a_0))\\
    &\quad+\sum \hat \mu_i \left(U(w, \hat a_i) - U(w, a_0)\right)
    \text{.}
    \end{aligned}
\]

Heuristically differentiating pointwise with respect to $w(y)$ and setting the derivative to zero gives
\[
    \frac{f(y \mid a_0)}{u'\bigl(w_0+w(y)\bigr)}
    =
    \lambda f(y \mid a_0)
    + \mu f_{a}(y\mid a_0)
    + \sum \hat \mu_i [f(y \mid a_0) - f(y \mid \hat a_i)]
    \text{.}
\]

The solution $w$ equals
\[
    w(y \mid \lambda, \mu, \boldsymbol{\hat \mu}, \boldsymbol{\hat a})
    :=
    L\biggl(
        \lambda + \mu s(y) + \sum \hat \mu_i \left(1 - \frac{f(y \mid \hat a_i)}{f(y \mid a_0)}\right)
    \biggr)-w_0
    \text{.}
\]

Define the Lagrange dual function as
\[
    \mathcal{D}(\lambda, \mu, \boldsymbol{\hat \mu}, \boldsymbol{\hat a})
    := \inf_{w}
    \mathcal{L}(w, \lambda, \mu, \boldsymbol{\hat \mu}, \boldsymbol{\hat a})
    =
    \mathcal{L}\bigl(w(\cdot \mid \lambda, \mu, \boldsymbol{\hat \mu}, \boldsymbol{\hat a}), \lambda, \mu, \boldsymbol{\hat \mu}, \boldsymbol{\hat a}\bigr)
    \text{.}
\]

For any parametric example, the dual can be computed efficiently with the analytical formulas from Section~\ref{sec:calculus}. It also has an analytic gradient by Danskin's envelope theorem. Moreover, given a grid of outputs $\mathbf{y}_{\mathrm{grid}}$, we can cache $f(y \mid a_0)$, $s(y)$, and $1 - {f(y \mid \hat a_i)}/{f(y \mid a_0)}$. This is enough to perform the numerical integrations needed for $\mathcal D$, so that $\mathcal D$ only involves matrix multiplications and applications of the wage link function $L$ and utility $u$.

This suggests Algorithm~\ref{alg:cost-minimization}:

\begin{algorithm}[H]
    \DontPrintSemicolon % Removes semicolons at end of lines for cleaner look
    \SetAlgoLined
    
    % Define keywords for functions to make them bold/distinct
    \SetKwFunction{FindBest}{FindBestDeviation}
    \SetKwFunction{MaximizeDual}{MaximizeDual}
    \SetKwFunction{InitGrid}{InitializeGrid}
    \SetKwFunction{UpdateCache}{UpdateCache}

    % Initialize variables
    % Using \tcp for comments ensures proper alignment
    $\boldsymbol{\hat a} \gets \emptyset$; \quad
    $\boldsymbol{\hat\mu} \gets \emptyset$; \quad
    $\lambda \gets \lambda_{\text{init}}$; \quad
    $\mu \gets \mu_{\text{init}}$\;
    
    $\mathbf{y}_{\mathrm{grid}} \gets \InitGrid()$\;

    \BlankLine
    \Repeat{$\mathtt{deviation\_gain} \le \mathtt{tolerance}$}{
        \UpdateCache{$\mathbf{y}_{\mathrm{grid}},\, a_0,\, \boldsymbol{\hat a}$} \tcp*[r]{For fast dual calls}
        
        $(\lambda,\, \mu,\, \boldsymbol{\hat\mu})
        \gets
        \MaximizeDual{$\mathcal{D}, \text{init}=\{\lambda, \mu, \boldsymbol{\hat\mu}\}$}$
        \tcp*[r]{use warm start}

        $w(\mathbf{y}_{\mathrm{grid}}) \gets w(\mathbf{y}_{\mathrm{grid}} \mid \lambda, \mu, \boldsymbol{\hat\mu}, \boldsymbol{\hat a})$\;

        $\mathtt{best\_deviation},\;\mathtt{deviation\_gain}
        \gets
        \FindBest{$w,\, a_0$}$\;

        \If{$\mathtt{deviation\_gain} > \mathtt{tolerance}$}{
            $\boldsymbol{\hat a} \gets \boldsymbol{\hat a} \cup \{\mathtt{best\_deviation}\}$\;
            $\boldsymbol{\hat\mu} \gets (\boldsymbol{\hat\mu}, 0)$ \tcp*[r]{Warm start update}
        }
    }

    \Return{$w$}

    \caption{Cost minimization via active set}
    \label{alg:cost-minimization}
\end{algorithm}

Algorithm~\ref{alg:cost-minimization} solves all the examples in the paper. In our benchmarks, each cost-minimization run takes on the order of 3 to 14 milliseconds (see supplementary material). Performance seems to be due to three factors. First, the algorithm often ends in the first iteration because of Theorem~\ref{thm:main}. Second, dual calls are fast because of the analytic formulas and caching. Third, dual maximization is fast because of low dimensionality, analytic gradients, global concavity of $\mathcal{D}$, and warm starts.

Algorithm~\ref{alg:cost-minimization} has one main limitation. For fixed $\boldsymbol{\hat a}$, the dual problem is concave. But in our implementation we search for global deviations over a continuous action space. Therefore, the active-set iterations are not guaranteed to terminate within a finite number of steps. We obtained good results both with the heuristic above and with the alternative of restricting attention to a finite grid of actions.

Due to this limitation, we implement a straightforward convex optimization solver. We discretize both the outcome and action spaces and formulate the problem using the utility contract $v(y)=u(w_0+w(y))$ parametrization from Appendix~\ref{subsec:appendix-notation}. In this parametrization, the objective is convex and the constraints are linear. We solve this discretized problem using \texttt{CVXPY} with the \texttt{Clarabel} solver. This approach is computationally more expensive, taking between 230 and 240 milliseconds in our benchmarks, because it solves for the contract $v$ at every point of the outcome grid. However, it provides a global optimality guarantee for the discretized problem. In our implementation, Algorithm~\ref{alg:cost-minimization} uses this convex solver as a fallback.

\begin{figure}[p]
    \centering
    \includegraphics[width=\textwidth]{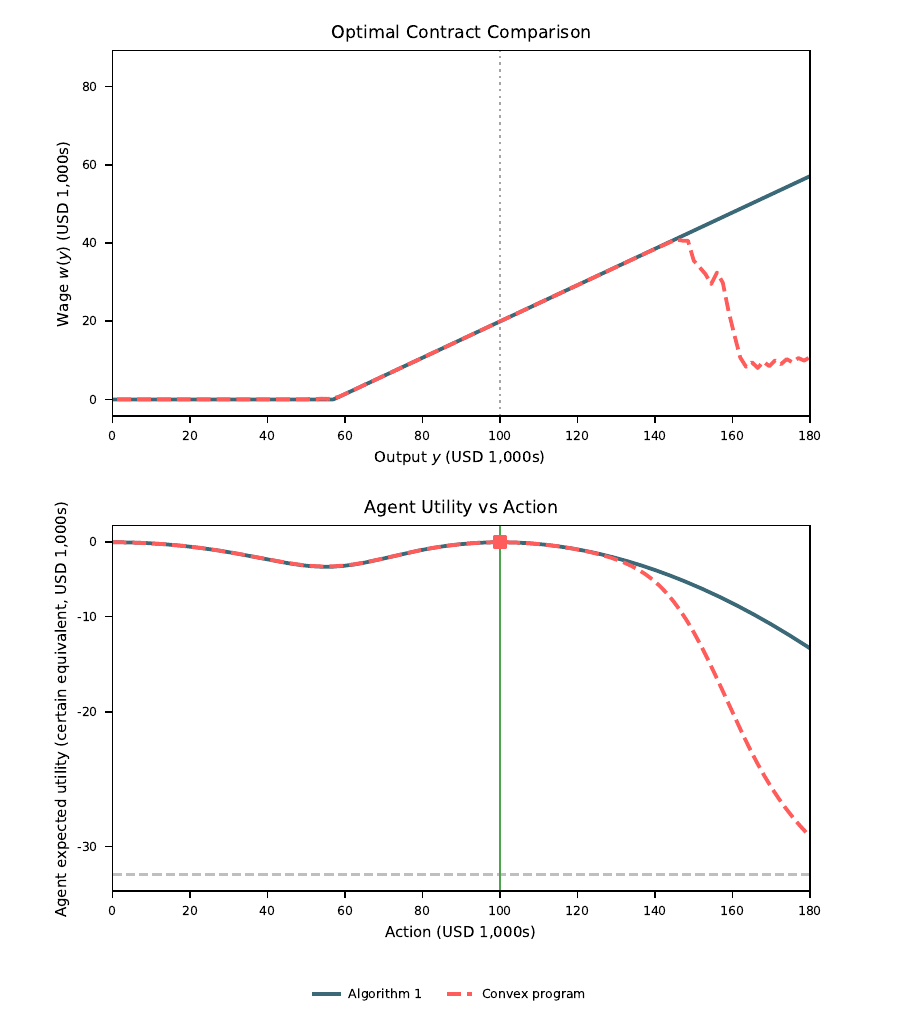}
    \captionsetup{font=footnotesize}
    \caption{Comparison of solvers for the cost-minimization problem.}
    \label{fig:convex-solver-stability}
    \caption*{\textit{Note:} Both solvers address the cost-minimization problem with fixed intended action $a_0=\$100{,}000$; the individual-rationality constraint does not bind. Top panel: optimal wage function $w(y)$. Bottom panel: agent's expected utility $U(w^*_{\bar U},a)$. Both solvers reach essentially the same expected wage cost. The convex-program solution is numerically unstable in the tails where the density $f(y \mid a_0)$ is small. Parameters are as in Figure~\ref{fig:gaussian-log-pp} with $\sigma=10$. The convex solver uses a grid of 201 points for outcomes and 200 points for actions.}
\end{figure}

The main drawback of the convex solver is that the numerical optimal contract $w(y)$ is numerically unstable at points where the density $f(y \mid a_0)$ is small. This is illustrated in Figure~\ref{fig:convex-solver-stability}. The figure shows that both solvers reach essentially the same objective value. In the middle of the distribution, they reach essentially the same optimal contract. However, in the tails of the distribution, the convex solver's optimal contract is numerically unstable.

\section{Discussion}
\label{sec:discussion}
\subsection{Numerical Experiments}
\label{sec:numerical-experiments}
Theorem \ref{thm:main} has two main shortcomings. First, it requires quantitative bounds on the safe score region. Thus, even if a parametric class of examples satisfies these conditions for some parameter values, the theorem is not guaranteed to hold for all parameter values.%
\footnote{%
This shortcoming appears in Corollary~\ref{cor:common-utilities}. The high-risk-aversion case in part (i) requires sufficiently low initial wealth. The log case in part (ii) requires that the score is unbounded below. This raises the question of whether the theorem commonly holds for reasonable levels of initial wealth in part (i), and whether it holds for common distributions with score bounded below in part (ii).}
Second, the theorem depends on ``sufficiently high'' reservation utility. Thus, the results are not guaranteed to work in economically meaningful examples. This limitation is similar to results like the central limit theorem, which is only useful because the ``large enough'' requirement tends to be reasonable in practical examples.

To understand how often the first-order approach holds or fails, we perform numerical experiments with 25 specifications. We consider outcome distributions Gaussian, Poisson, exponential, gamma, geometric, Bernoulli, binomial, and Student $t$. The $t$ distribution is included as an adversarial case, as in all $t$ distribution examples we consider the safe score region is empty, thus not satisfying the theorem's assumptions. In the Gaussian case, we vary some key parameters. For risk aversion we try CRRA and CARA with relative risk aversion from $0.25$ to $2.0$. We vary the amount of noise, initial wealth, and the cost of effort. Parameters are chosen to represent a task that produces revenue to the principal of the order of \$100,000, but with meaningful effort cost and uncertainty so that agency is costly. Effort cost is quadratic. Appendix \ref{sec:foa-numerical-experiments} gives the exact specifications.

Figure~\ref{fig:foa-principal-summary} summarizes results for the principal's problem and Supplementary Appendix~\ref{sec:foa-numerical-experiments} gives the qualitatively similar results for the cost-minimization problem. The results are as follows, setting aside for now the adversarial $t$ distribution. The first-order approach holds for almost all reservation wages. We plot as reference points the monopsony wage and the competitive wage. These correspond to the reservation utilities agents receive when the principal has no individual rationality constraint (monopsony), or when the principal has zero profits (competitive). In all cases, the first-order approach holds for every reservation wage above a threshold that is at most slightly higher than the monopsony wage. In many cases, the red region where the first-order approach fails is so small it is difficult to perceive visually.

These results are at odds with the standard view in the literature. The standard view is that the first-order approach often fails because none of the examples satisfy existing sufficient conditions. Moreover, there is no standard reason for the first-order approach to fail in the cases of low or high reservation utility. In contrast, in the experiments the first-order approach holds for almost all plausible reservation utilities, with the failures being concentrated at low reservation utility.

The main shortcoming of the experiments is that, due to space constraints, we included only a finite set of examples. To make the analysis transparent, we provide a \href{https://colab.research.google.com/github/eduardomazevedo/moralhazard/blob/main/examples/colab_demo.ipynb}{Google Colab notebook} and \href{https://eduardomazevedo.github.io/azevedo-wolff-2025-site/}{interactive demonstration}, so readers can easily explore their own examples.

\subsection{Principal's Problem versus Cost Minimization}
\label{sec:principal-versus-cost-minimization}

All of our theoretical results are stated for the cost-minimization problem, where we keep the intended action $a_0$ fixed. But in applications we are often interested in the principal's problem, where the principal optimizes over the intended action $a_0$. This leads to the question of when the first-order approach is valid for the principal's problem, both in practical examples and in extensions of our theoretical results.

In terms of numerical experiments, we illustrated our results both in the principal's problem (Figures~\ref{fig:gaussian-log-pp}, \ref{fig:poisson-log-pp}, \ref{fig:gaussian-cara-pp}, \ref{fig:foa-principal-summary}, and \ref{fig:t-log-pp}) and in cost-minimization problems (Figures~\ref{fig:pf}, \ref{fig:gaussian-log-cm}, and \ref{fig:foa-fixed-action-summary}). The experiments suggest that the first-order approach is broadly valid in both cases, except at low reservation utility and in the adversarial Student-$t$ examples, which have an empty safe score region.

Interestingly, in the experiments, the thresholds for validity of the first-order approach are often lower in the principal's problem than in the cost-minimization problem (compare Figures~\ref{fig:foa-principal-summary} and \ref{fig:foa-fixed-action-summary}). The intuition is that, with high reservation utility, the principal's optimal action is often close to the lowest possible action. We chose the intended actions in the cost-minimization examples to be close to the first-best actions. These tend to be higher than the principal's optimal actions, and the differences tend to grow as reservation utility increases.

This intuition is important for theoretical extensions. Under mild assumptions, increasing $\bar U$ leads the principal's optimal action to approach the lowest possible action. Thus, one could give reasonable definitions and assumptions under which the first-order approach in the principal's problem ``is valid'' at high reservation utility, but only because the principal's optimal intended action is being pushed down to the boundary.%
\footnote{For example, one can make assumptions under which, for sufficiently high reservation utility, the lowest possible action is trivially the only solution to both the principal's problem and a particular relaxed version in which the local incentive constraint at the boundary is $U_a(w,a_0)\leq0$. Suppose $\mathcal A=[0,\bar a]$, $c(0)=0$, $c'(0)>0$, and $\Revenue(a)-\Revenue(0)\leq \bar\beta a$ for some $\bar\beta<\infty$. Let $W_R:=u^{-1}(\bar U)$. For any contract satisfying individual rationality at action $a$, Jensen's inequality implies
\[
 C(w,a)\geq u^{-1}\bigl(\bar U+c(a)\bigr)-w_0.
\]
Convexity of $c$ and of $u^{-1}$ therefore gives
\[
 u^{-1}\bigl(\bar U+c(a)\bigr)-u^{-1}(\bar U)
 \geq \frac{c'(0)}{u'(W_R)}a.
\]

Because $u'(W_R)\to0$ as $W_R\to\infty$, this additional compensation cost exceeds $\bar\beta a$ for sufficiently high $\bar U$. At $a_0=0$, the constant contract $w(y)=W_R-w_0$ satisfies individual rationality, globally implements $0$, satisfies the boundary local incentive constraint, and attains the lower bound. Hence, for sufficiently high $\bar U$, it solves the cost-minimization problem with intended action $0$, the principal's problem, and the boundary-relaxed principal's problem. The conclusion is driven entirely by the boundary action, rather than by any substantive first-order-approach argument.}

There are three main reasons why our results are stated in terms of the cost-minimization problem. First, fixing the intended action avoids results driven trivially by the principal's optimal action being pushed to the boundary, as in the footnote above. Such results can make the original and relaxed principal's problems coincide, but they do not clarify the mechanism behind Theorem~\ref{thm:main}, which makes the first-order approach valid even in the cost-minimization problem. Second, we do not have to make assumptions on the revenue function $\Revenue ()$, and output $y$ matters only as a statistical signal, so the analysis is invariant to one-to-one, action-independent relabelings. Third, the analysis of the cost-minimization problem is simpler \textcite{grossman1983}.

We see three promising directions for extending our results to the principal's problem. First, the large-$\lambda$ concavity argument in Proposition~\ref{prop:concave} appears capable of being made uniform in $a_0$. Uniform versions of the theorem's bounds and implementability conditions could therefore yield validity simultaneously for every relevant interior intended action. Second, one could combine such an argument with a result localizing the principal's candidate optimal intended actions near the lowest possible action, using logic similar to the footnote above while allowing the optimum to remain interior. It might then suffice for the conditions of Theorem~\ref{thm:main} to hold uniformly only on a small range of intended actions near the lower boundary, rather than on all of $\mathcal A$. The third direction is to consider a parametrization of the $\Revenue ()$ function that depends on $\bar U$, so that the optimal action in the relaxed problem stays fixed. This is similar to the approach developed by \textcite{chade2020highstakes}. In the interest of space, we leave these extensions to future work.

\subsection{Importance of Assumptions and Counterexamples}

We now consider the importance of the main assumptions.

\label{sec:counter-examples}
\medskip\noindent\textit{\textbf{Safe Score Region.}}
A key assumption of Theorem~\ref{thm:main} is the existence of a sufficiently large safe low-score region $\mathcal Q_{\mathrm{safe}}$.  The region must be large enough for the required conditions on $\mathrm{Curv}_{\mathrm{safe}}$ and $\mathrm{Cap}_{\mathrm{safe}}$ to hold.

To see the importance of this region, consider the Student-$t$ distribution.  The $t$ distribution resembles a Gaussian but has fatter tails.  Consequently, its score resembles the Gaussian score near the mean but approaches zero in both tails (Table~\ref{tab:dists}).  The score is not monotone and there may be no nonempty safe low-score region.%
\footnote{Let $y^*:=a_0-\sigma\sqrt{d}$, where the intended-action score attains its minimum; every nonempty score sublevel set contains $y^*$.  Moreover,
\[
 \frac{f_{aa}(y\mid a)}{f(y\mid a)}
 =\frac{(d+1)[(d+2)(y-a)^2-d\sigma^2]}
 {[d\sigma^2+(y-a)^2]^2}.
\]
If $y^*\in\mathcal A$, evaluating at $a=y^*$ gives
$f_{aa}(y^*\mid y^*)/f(y^*\mid y^*)=-(d+1)/(d\sigma^2)<0$, independently of the parameter values.}

\begin{figure}[p]
    \centering
    \includegraphics[width=\textwidth]{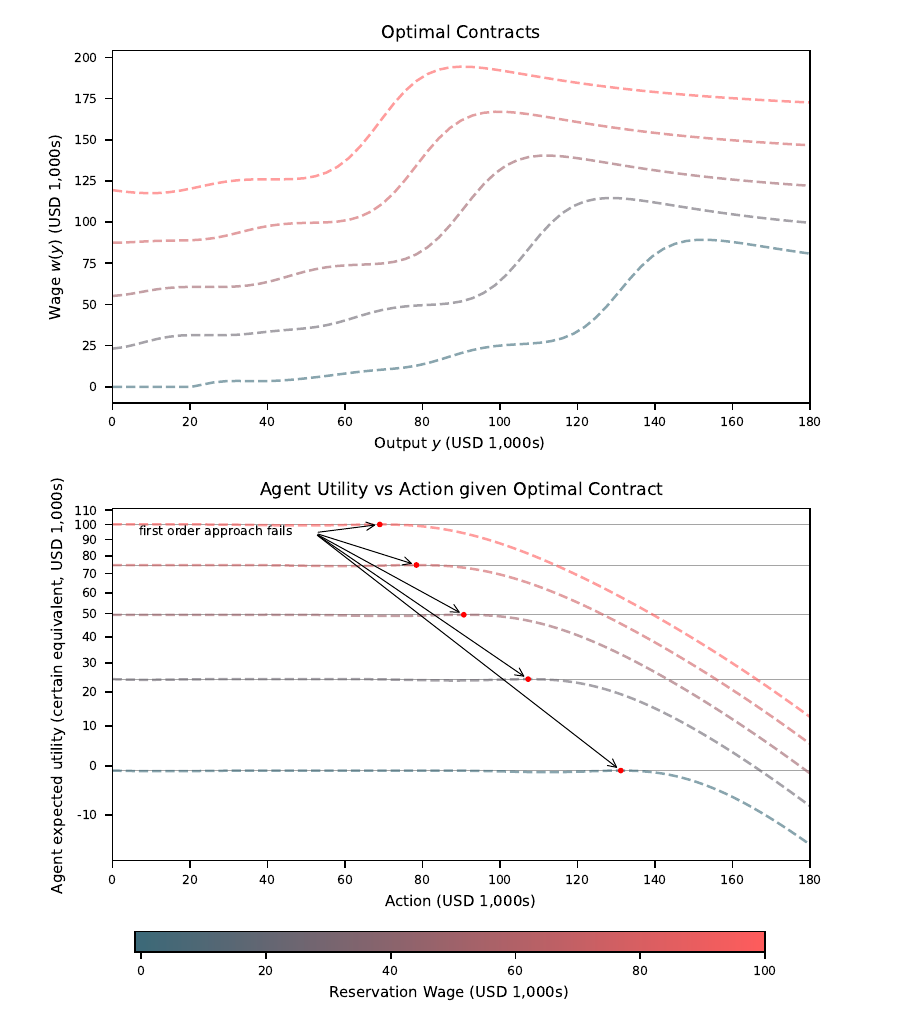}
    \captionsetup{font=footnotesize} % Makes the note footnote-sized
    \caption{Optimal contracts in the principal's problem with Student-$t$ output and log utility.}
    \label{fig:t-log-pp}
    \caption*{\textit{Note:} Top panel: optimal wage function $w(y)$. Bottom panel: agent's expected utility $U(w^*_{\bar U},a)$ under the displayed contract. For each reservation utility, the displayed contract and recommended action solve the principal's problem. Colors represent reservation utility. Dashed curves indicate that the first-order approach is invalid for the corresponding recommended action and reservation utility. Dots in the bottom panel indicate the recommended action. The thin horizontal line indicates indifference between local maxima. Output follows a Student-$t$ distribution with mean $a$, scale parameter $\sigma=20.0$, and degrees of freedom $d=1.15$. Initial wealth is $50$ (output and wealth are in thousands of dollars), and the cost function is $c(a) = a^2 / 30000$.}
\end{figure}

This results in common failures of the first-order approach (Figures~\ref{fig:t-log-pp} and \ref{fig:foa-principal-summary}). We stress, however, that our conditions are sufficient but far from tight. In Figure~\ref{fig:foa-principal-summary}, all the non-$t$-distribution examples have nonempty safe score regions, but not all of them satisfy the quantitative capacity or curvature conditions. Nevertheless, we observe that the first-order approach is valid for most reservation utilities in these examples.

% Student-t - log utility figure

\medskip\noindent\textit{\textbf{Sufficient Risk Aversion.}}
Risk aversion is an important assumption. In the risk-neutral case, there are simple counterexamples where the first-order approach fails even when the distributional conditions of Theorem~\ref{thm:main} are satisfied (Supplementary Appendix~\ref{sec:risk-neutral-counterexample}).

For practical purposes, the first-order approach can be valid with considerably lower risk aversion than the sufficient conditions in Theorem~\ref{thm:main} require. In the CRRA case, Theorem~\ref{thm:main} covers coefficients of relative risk aversion of $1$ or more. Supplementary Appendix~\ref{sec:asymptotic-affinity}, Corollary~\ref{cor:asymptotic-affinity-foa}, gives a sufficient condition that works for every positive coefficient of relative risk aversion below $1$, at the cost of stronger distributional assumptions. The main additional assumption is concavity of the expected score, $a\mapsto\E_a[s(Y)]$. Corollary~\ref{cor:gaussian-crra-extension} shows that this extension applies to Gaussian, Poisson, exponential, gamma, geometric, Bernoulli, and binomial output.

\medskip\noindent\textit{\textbf{Limited liability.}}
Limited liability makes it more difficult to guarantee validity of the first-order approach, due to the kink in optimal contracts. We impose limited liability because it is crucial for existence of an optimal contract. The intuition is clear from the Lagrangian's first-order condition in equation~\eqref{eq:foc}. When the right-hand side is negative, it is optimal to give the agent the lowest possible $w(y)$. For a solution to exist, we often need limited liability and $u(w_0) > - \infty$. When the score is unbounded below, the right-hand side is negative for any $\mu >0$; thus, existence problems are extremely common without limited liability. These existence issues have been well known since \textcite{mirrlees1999}.

\subsection{Literature Review}
\label{sec:literature-discussion}
\medskip\noindent\textit{\textbf{First-Order Approach Literature.}}
There are three generations of contributions to the first-order approach. The first generation was initiated by \textcite{mirrlees1999} (circulated in 1975), who showed that the first-order approach is not generally valid. Prior to \textcite{mirrlees1999}, the first-order approach was simply assumed with little justification. Subsequent authors recognized that they had to assume the validity of the first-order approach \parencite{holmstrom1978incentives}, and developed results that do not depend on the first-order approach \parencite{grossman1983,araujo2001general}.

The seminal papers in the second generation are \textcite{rogerson1985} and \textcite{jewitt1988justifying}. They provided sufficient conditions for the first-order approach. Rogerson's conditions are well known to be restrictive, ruling out many natural distributions, including the Gaussian. Jewitt's conditions are more general and include many interesting cases. His theorem, however, does not impose a lower bound on payments. He considers a solution to the relaxed problem that satisfies the interior pointwise first-order condition for wages globally. His concavity argument therefore does not apply when limited liability binds. The limited liability kink destroys the global concavity used in the proof. \textcite{chaigneau2022should} explain this issue in detail and survey the related literature.

Jewitt's theorem establishes validity conditional on a regular solution to the relaxed problem but does not guarantee that such a solution exists. For Gaussian output and log utility without limited liability, Jewitt's stated shape conditions are satisfied, but the relaxed cost-minimization problem has no solution without limited liability. Whenever $\lambda + \mu s(y)<0$ the first-order condition would ask for negative $1/u'(w(y)+w_0)$. Jewitt's theorem therefore does not apply to an attained optimum in this example. The second-generation papers have been influential, and much of the theoretical and applied work to this day assumes these conditions.

Our own results are inspired by \textcite{jewitt1988justifying}, and we build on his approach. We impose similar conditions in the portion of the optimal contract where limited liability does not bind, but allow for limited liability.%
\footnote{Jewitt defines $\omega(z)=u((u')^{-1}(1/z))$, which is our base utility link function $\ell_0(z)$. His assumption that $\omega$ is concave is exactly Condition~\eqref{condition:U1}. His assumptions on output are stronger than ours: among other conditions, he assumes that expected output is nondecreasing and concave in action. His concavity-preservation and monotone-concave-score assumptions also imply that $a\mapsto\E_a[s(Y)]$ is concave, where $s$ is the score at the intended action. We impose this latter condition in the CRRA extension in Supplementary Appendix~\ref{sec:asymptotic-affinity}, but not in our main theorem.}
The proof of Theorem~\ref{thm:main} shows that, with high reservation utility, the limited liability constraint binds with low probability. We then show, along lines similar to \textcite{jewitt1988justifying}, that the agent's problem becomes concave.

There is a third generation of work extending Jewitt's approach to multidimensional actions and richer information structures \parencite{conlon2009two,jung2015information,chade2020no,chaigneau2022should,jung2024proxy}. We believe our methods can be adapted to such environments, but we do not pursue that here.

At first glance, our Theorem~\ref{thm:main} appears to contradict the standard view in the literature that the first-order approach requires stringent conditions. In fact, the two are perfectly consistent. The reason is that the literature has looked for conditions under which the first-order approach is valid \emph{for every reservation utility}. Our examples show that indeed any such condition must rule out basic examples like the Gaussian-log utility example from Figure~\ref{fig:gaussian-log-pp} -- and in fact all the numerical examples in this paper. Our contribution is to show that \emph{if the reservation utility is sufficiently high}, the first-order approach is valid.

\medskip\noindent\textit{\textbf{Linear Contracts and Option Contracts.}}
A long literature seeks conditions under which optimal contracts are simple, such as linear contracts. The classic paper by \textcite{holmstrom1987aggregation} rationalizes linear contracts with CARA utility in a continuous-time dynamic model with a Brownian output process. \textcite{innes1990limited} rationalizes piecewise linear contracts with a risk-neutral agent, limited liability, and a monotonicity restriction on the wage function. Notable contributions include \textcite{chassang2013calibrated} and \textcite{carroll2015robustness} who rationalize linear contracts with risk-neutral agents and a robust maxmin notion of optimality.

In contrast, we derive piecewise linear contracts under log utility and linear score. Our contracts are option contracts, with a payoff of zero for low outcomes and linearly increasing payment past a strike price. In some examples, our optimal contracts are linear in most of the support of the distribution of output. In some examples with a finite lower bound on the support, our optimal contracts are linear in the entire support.

The most closely related work is the contemporaneous paper \textcite{opp2025moral}. He independently derived the linearity result from Remark~\ref{rem:linear-contracts}. \textcite{opp2025moral} proves that the first-order approach is valid with an exponential distribution for some parameter values. As far as we know, \textcite{opp2025moral} is the first to derive piecewise linear contracts under a broad set of conditions in standard moral hazard with risk aversion. \textcite{opp2025moral} includes a complete analysis of the exponential distribution case. He characterizes optimal contracts with and without limited liability and when the optimal contract exists without limited liability. In particular, he shows that the optimal contract with limited liability is globally linear for low intended action but has a kink for high intended action. Except for the basic linearity result, these results are not covered by our paper.

In the accounting literature, we know of two papers that showed that (potentially piecewise) linear contracts are optimal under certain assumptions, including the validity of the first-order approach. The earliest is \textcite{hemmer1999introducing}. They assume a number of Jewitt's conditions, and log utility with no limited liability. Their Corollary 2, part 2, then shows that the optimal contract is globally linear. They prove that contracts are globally linear, whereas we prove that contracts are piecewise linear. While their result is stronger, optimal contracts often do not exist in their setting (Section~\ref{sec:counter-examples}). The excellent graduate textbook by \textcite{christensen2005economics}, Chapter 19, p. 154, also notes that piecewise linear contracts are optimal assuming that the first-order approach is valid. Thus, they get to the generally correct form of piecewise linear contracts as opposed to the globally linear result. But they assume the first-order approach, and do not prove existence. Thus, \textcite{opp2025moral} and our paper are the first to derive piecewise linear contracts under a broad set of conditions, rather than assuming the first-order approach. But \textcite{hemmer1999introducing} were the first to notice linearity of the base wage link function $L_0$, to the best of our knowledge. \textcite{chaigneau2015changes,chaigneau2017prudence} also attribute this observation to \textcite{hemmer1999introducing}.

Another closely related paper is \textcite{chaigneau2022should}. They consider the Gaussian-log utility case with limited liability, and assume the validity of the first-order approach. Their Lemma 2 shows that option contracts are optimal, as in our Remark~\ref{rem:linear-contracts}. This is the first result we know justifying option contracts in the sense of zero payment for low outcomes and linearly increasing payment past a strike price. \citeauthor{holmstrom1978incentives}'s (\citeyear{holmstrom1978incentives}) thesis considers the case of log utility with an upper and lower bound on the wage function. He assumes the first-order approach and shows that the optimal contract is linear within those bounds.

\medskip\noindent\textit{\textbf{Solution of the Relaxed Cost-Minimization Problem.}}
The solution to the relaxed cost-minimization problem in Proposition~\ref{prop:relaxed-optimal-contract} is a trivial variation of the standard solution using the first-order approach. The case of limited liability is considered in \citeauthor{holmstrom1978incentives} (\citeyear{holmstrom1978incentives}, \citeyear{holmstrom1979moral}). \textcite{jewitt2008moral} consider an even more general case, with lower and upper bounds that depend on output. Their elegant Definition 1 is basically the definition of a canonical contract, and their Proposition 1 is basically the same as Proposition~\ref{prop:relaxed-optimal-contract}. We provide a proof of Proposition~\ref{prop:relaxed-optimal-contract} because our assumptions are different. \textcite{jewitt2008moral} assume the support of $y$ is a compact interval, the convexity of the distribution function condition, and bounded score. They argue that the bounded score is important for existence (p. 62), although in our setting solutions exist even with unbounded score. \textcite{moroni2014existence} corrects a mistake in \textcite{jewitt2008moral}, showing the need for an assumption similar to utility being bounded below, which we make. Our existence proofs are different. \textcite{jewitt2008moral} prove that given a multiplier $\mu$ there exists a single multiplier $\lambda$ as a function of $\mu$ where the canonical contract satisfies the individual rationality constraint. They then use the intermediate value theorem to show that there exists a $\mu$ where the canonical contract satisfies all constraints. We instead prove (1) that the Lagrangian has a unique minimum that is a canonical contract (Lemma~\ref{lem:lagrangian}), (2) that the Pareto problem has a solution that minimizes the Lagrangian (Lemma~\ref{lem:pareto-problem}), and (3) that the Pareto problem solution solves the relaxed cost-minimization problem (proof of Proposition~\ref{prop:relaxed-optimal-contract}, part 2). While we include the proof of Proposition~\ref{prop:relaxed-optimal-contract} for completeness, advanced readers may take it as trivial given the \citeauthor{holmstrom1978incentives} (\citeyear{holmstrom1978incentives}, \citeyear{holmstrom1979moral}), \textcite{jewitt2008moral}, and \textcite{moroni2014existence} results.

\medskip\noindent\textit{\textbf{High Reservation Utility.}}
In an influential paper, \textcite{chade2020highstakes} consider the principal-agent problem in the case where reservation utility converges to infinity. This paper is one of our inspirations in considering the case of high reservation utility, although it deals with completely different issues. They assume the first-order approach is valid (their Assumption 4), and give no results on validity of the first-order approach. Their main result is that the wage cost to the principal is asymptotically the same as if effort was observable, so that agency costs are asymptotically small. This is a strong result that depends on detailed assumptions that rule out many of our examples. As they note, ``key assumptions driving our results are on the agent's utility for income'' \ldots{} ``they are not without bite: log utility is excluded, and indeed the convergence results fail in this setting.''

The mechanism behind their result is transparent with CRRA utility $u(W)=W^{1-\gamma}/(1-\gamma)$. Let $W_R$ be the wealth that provides the agent her reservation utility. Under observable effort, compensating the agent for an effort cost $c$ requires an additional payment asymptotic to $c/u'(W_R)=cW_R^\gamma$. Providing incentives also requires variation in the agent's utility across outcomes. Under their assumptions, this variation remains bounded as $W_R$ grows. At wealth $W_R$, generating a fixed amount of utility variation requires wage variation of order $1/u'(W_R)=W_R^\gamma$, whose variance is therefore of order $W_R^{2\gamma}$. The additional average wage needed to compensate the agent for bearing a small risk is proportional to absolute risk aversion times the variance of the risk. Since CRRA absolute risk aversion is $\gamma/W_R$, the agency cost is of order $W_R^{2\gamma-1}$. Consequently, its ratio to the first-best cost of effort is of order $W_R^{\gamma-1}$, which converges to zero when $\gamma<1$. Moral hazard therefore affects costs only at a lower order than the first-best cost of compensating effort. For log utility, corresponding to $\gamma=1$, the two costs have the same order, and their conclusion can fail. While we draw on their insight of considering high reservation utility, this mechanism clarifies the differences and our marginal contribution. First, they assume the first-order approach is valid and give no results on validity of the first-order approach. Second, they make different and strong assumptions that rule out much of our analysis, in particular assuming low risk aversion rather than large risk aversion. Third, our results broadly hold in many settings where their asymptotics do not. For example, their results show that agency costs are asymptotically small compared to reservation wages. In all of the numerical examples in this paper, agency costs are significant; nevertheless, we see that the first-order approach is broadly valid.

\textcite{castro2024disentangling} assumes that the principal's expected wage cost is convex in the intended action, which they motivate with a result by \textcite{chade2020highstakes} that this is true with high stakes. But \textcite{castro2024disentangling} does not otherwise use high reservation utility. They assume that the first-order approach is valid (their Assumption 1). Neither paper assumes limited liability, and \textcite{castro2024disentangling} assumes that utility is unbounded below but uses a bounded below utility in their numerical results. \textcite{chade2026initiative} consider a moral hazard problem with a set of actions in an interval plus one non-ordered action. Their analysis is particularly tractable with CRRA coefficient $\gamma=1/2$, because the base utility link function $\ell_0$ is linear. Once reservation utility is high enough that limited liability does not bind, the relaxed cost-minimizing contract in utility units is therefore affine in the relevant likelihood ratios. To write their formula in our notation, let $r(y):=f^s(y)/f(y\mid a_0)$ be the likelihood ratio of the non-ordered action to the intended action, and let expectations be under $f(\cdot\mid a_0)$. Writing $v(y):=u(w_0+w(y))$, their optimal contract is $v(y)=\lambda+\mu s(y)-\eta r(y)$. Its multipliers satisfy the linear system
\[
    \lambda-\eta=\bar U+c(a_0),\qquad
    \mu\mathbb E[s^2]-\eta\mathbb E[sr]=c'(a_0),\qquad
    \lambda+\mu\mathbb E[sr]-\eta\mathbb E[r^2]\leq\bar U,
\]
where $\eta\geq0$ and $\eta\{\bar U-[\lambda+\mu\mathbb E[sr]-\eta\mathbb E[r^2]]\}=0$. Building on \textcite{chade2020highstakes}, they show that, for a broader class of utility functions at high stakes, the relaxed contract in utility units converges to this $\gamma=1/2$ formula. They give a high-stakes result on validity of the first-order approach. They show that if the $\gamma=1/2$ limiting contract globally implements the intended action, with strict incentives, then the contracts for the other utility functions also do so at sufficiently high reservation utility. They also provide conditions under which global incentive compatibility in the $\gamma=1/2$ relaxed optimal contract holds. Our marginal contribution is to provide results that do not require the first-order approach to hold in the $\gamma=1/2$ benchmark. Instead, we give broadly applicable primitive conditions under which high reservation utility makes the first-order approach valid, including CRRA coefficients $\gamma\geq1$, which are not covered by their high-stakes analysis.

\medskip\noindent\textit{\textbf{Numerical Analysis.}}
We now review existing approaches to numerically solve the moral hazard problem without the first-order approach. The review suggests that Algorithm~\ref{alg:cost-minimization} offers a significant improvement. We are, to the best of our knowledge, the first to numerically solve the moral hazard problem for various realistic distributions without the first-order approach.

The literature proposes four general approaches to numerically solve the moral hazard problem without the first-order approach. The first approach is to include additional no-jump constraints in the relaxed cost-minimization problem. This was introduced in the 1970s by \textcite{mirrlees1999} and improved by \textcite{araujo2001general}. The main limitation is that there is no general method to decide which constraints to add (see \textcite{ke2018general}, p. 1426). Motivated by this, \textcite{ke2018general} introduce a sandwich method in which the additional constraint is added adversarially. This method works well in some special cases that are analytically tractable, such as relative risk aversion with coefficient $1/2$ where the wage link function is piecewise linear.

The two most popular approaches start by discretizing the set of actions, so that there is only a finite number of global incentive compatibility constraints (following \textcite{grossman1983}). The linear programming approach maximizes over a joint distribution of actions, output levels, and utility from wages, all of which have to be in a discrete grid \parencite{prescott1999primer, prescott2004computing, su2007computation}. While this has the advantage of making the principal's problem a linear program, the main difficulty is the curse of dimensionality: the number of variables is the product of the sizes of the three grids. To address this, \textcite{su2007computation} introduced the mathematical program with equilibrium constraints approach. This approach maximizes over the contract and a distribution over actions. So the number of variables is the number of possible output values plus the number of possible actions.

We know of two papers that apply these approaches to realistic distributions. \textcite{prescott2004computing} considers a Gaussian distribution with an improvement on the linear programming approach by using the Dantzig-Wolfe decomposition algorithm. \textcite{armstrong2007stock} consider a lognormal distribution to study executive compensation and use a mathematical program with equilibrium constraints approach. Both papers are far ahead of their time, as their numerical solutions display the key features of optimal contracts. They also report the wall time needed to solve the principal's problem. \textcite{prescott2004computing} reports 51 minutes with 100 actions and 50 output levels. \textcite{armstrong2007stock} reports that each solution takes ``several hours'' with 101 actions and 501 output levels. In our supplementary material benchmark table, Algorithm~\ref{alg:cost-minimization} takes 93 to 356 milliseconds to solve similar Gaussian problems (without the need to discretize either actions or output levels -- except when evaluating integrals). Naturally, these are not apples-to-apples comparisons due to improvements in hardware and implementation details. But the large performance gap suggests that Algorithm~\ref{alg:cost-minimization} is substantially faster. This is consistent with Algorithm~\ref{alg:cost-minimization} being significantly faster than the convex solver (see supplementary material). This is intuitive because Algorithm~\ref{alg:cost-minimization} uses the analytic formulas from Tables~\ref{tab:utility_functions} and \ref{tab:dists} and solves lower-dimensional problems.

The fourth approach in the literature is to approximate the agent's utility as a function of actions with a ratio of polynomials \parencite{renner2015polynomial}. \textcite{renner2015polynomial} then use results from polynomial optimization to show that the principal's problem can be reduced to a nonlinear program. Crucially, their method can give global optimality guarantees. They solve the lognormal example from \textcite{armstrong2007stock} in the case of a restricted contract space, with a fixed payment, equity, and options. They generously shared their code. Our benchmarks suggest that their suggested nonlinear program takes about 100 seconds in SciPy's SLSQP solver.

Our approach is influenced by and builds upon this literature. Algorithm~\ref{alg:cost-minimization} is most related to the no-jump approach. We simply use a reasonable heuristic to iteratively add constraints, while taking advantage of Lagrange duality, parametric assumptions for fast dual calculations, analytic gradients, caching, and warm starts. Algorithm~\ref{alg:cost-minimization} has the disadvantage of not giving a global optimality guarantee, and is thus complementary to the LP and polynomial approaches. For example, one could use our algorithm to obtain a solution candidate and check it with the LP or polynomial approaches. Our convex programming approach is a natural modern extension of the linear programming and mathematical programming with equilibrium constraints approaches. Since the 2010s, efficient convex solvers that are simple for nonexperts to use have become increasingly available and are now standard tools for solving these types of problems in engineering. Our convex programming approach simply applies these new tools to the standard discretization approach from the literature going back to \textcite{grossman1983}.

\section{Conclusion}
\label{sec:conclusion}
The standard view in the literature is that there are widespread failures of the first-order approach and that, due to these global deviations, the moral hazard problem is often intractable. Moreover, the optimal contracts are complex and unrealistic. We make two marginal contributions. First, we find that the first-order approach is valid in several applied examples when reservation utility is not too low. In particular, optimal contracts are often tractable, simple, and match broad features of real contracts like option contracts. Second, even when the first-order approach fails, we show that the optimal contract can be computed trivially. We hope that these results will broaden the applicability of contract theory.

% Print bibliography
\printbibliography

\appendix
\newpage
\begin{center}
{\huge Appendix}{\huge\par}
\par\end{center}

\section{Proofs\label{sec:appendix-proofs}}
\newcommand{\Capq}{\mathrm{Cap}}
\newcommand{\Curvq}{\mathrm{Curv}}

\subsection{Notation and Utility Parametrization}
\label{subsec:appendix-notation}

In the appendix we use the following additional notation. Let
\[
    \underline u:=u(w_0),\qquad
    \bar u:=\lim_{W\to\infty}u(W),\qquad
    \Delta u:=\bar u-\underline u,
\]
and $k:=u^{-1}$.

It will be convenient to parametrize contracts in utility units because this
makes the relaxed problem a convex program with linear constraints.  Given a
payment contract $w$, define its \textbf{utility contract} by
$v(y):=u(w_0+w(y))$.  Conversely, a utility contract $v$ corresponds to the
payment contract $w(y)=k(v(y))-w_0$.  Define
\begin{align*}
    \widehat U(v,a)&:=\int v(y)f(y\mid a)\,d\nu(y)-c(a),\\
    \widehat C(v,a)&:=\int [k(v(y))-w_0]f(y\mid a)\,d\nu(y).
\end{align*}
For corresponding utility and payment contracts,
\begin{equation}
    \widehat U(v,a)=U(w,a),
    \qquad
    \widehat C(v,a)=C(w,a).
    \label{eq:utility-payment-equivalence}
\end{equation}
Henceforth, throughout the proofs appendix, we work in utility units and refer
to utility contracts simply as contracts.

Local incentive compatibility in utility units is
\begin{equation}
    \widehat U_a(v,a_0)
    =0.
    \tag{$\widehat{\mathrm{LIC}}$}\label{LIC_hat}
\end{equation}
The Lagrangian in utility units is
\begin{equation}
    \widehat{\mathcal L}(v,\lambda,\mu)
    :=\widehat C(v,a_0)
      +\lambda[\bar U-\widehat U(v,a_0)]
      -\mu\widehat U_a(v,a_0).
    \label{eq:utility-lagrangian}
\end{equation}

Define the feasible-contract set in utility units by
\[
    \mathcal V
    :=\left\{v:\mathcal Y\to[\underline u,\bar u):
    k(v)-w_0\in\mathcal W\right\}.
\]
$\mathcal V$ is convex: differentiation is preserved by convex combinations, and
convexity of $k$ preserves finiteness of expected compensation cost.

Define the base utility link function and limited-liability utility link function by
\begin{eqnarray*}
    \ell_0(z)&:=&u(L_0(z))=(k')^{-1}(z),\\
    \ell(z)&:=&\ell_0(z\vee m_0).
\end{eqnarray*}
The body uses the wage link function $L$, whereas the proofs below use only the
utility link functions $\ell_0$ and $\ell$.  To reduce terminology, throughout
the remainder of this appendix ``link function'' means utility link function
unless explicitly stated otherwise.
The utility representation of the canonical payment contract is
\[
    v_{\lambda,\mu}(y)
    :=u\bigl(w_0+w_{\lambda,\mu}(y)\bigr)
    =\ell(\lambda+\mu s(y)).
\]

\subsection{Utility Conditions in Link-Function Form}
\label{subsec:appendix-link-conditions}

Our utility conditions imply the following properties of the base link function $\ell_0$.

\begin{remark}[Concavity of the link function]
\label{rem:link-concavity}
If Condition~\eqref{condition:U1} holds, then $\ell_0$ is concave on
$[m_0,\infty)$.
\end{remark}

\begin{proof}
For $z>m_0$, let $W=W(z)$ be determined by
$z=1/u'(W)$, so that $\ell_0(z)=u(W(z))$.  Suppressing the argument
$W(z)$ below, differentiation gives
\[
    W'(z)=-\frac{u'^2}{u''}>0.
\]
Hence
\[
    \ell_0'(z)=u'W'(z)
    =-\frac{u'^3}{u''}>0.
\]
Differentiating again,
\begin{eqnarray*}
    \ell_0''(z)
    &=&W'(z)\frac{d}{dW}
       \left(-\frac{u'^3}{u''}\right)\\
    &=&W'(z)u'^2
       \left[\frac{u'u'''}{u''^2}-3\right]\\
    &=&W'(z)u'^2
       \left[\frac{P}{A}-3\right].
\end{eqnarray*}
Condition~\eqref{condition:U1} implies $P>0$ and $P/A\leq3$.
Therefore $\ell_0''(z)\leq0$.
\end{proof}

\begin{remark}[Asymptotic flattening of the link function]
\label{rem:link-flattening}
Condition~\eqref{condition:U2} is equivalent to
\[
    \lim_{z\to\infty}z\ell_0'(z)=0.
\]
\end{remark}

\begin{proof}
For $z>m_0$, let $W=W(z)$ be determined by
$z=1/u'(W)$.  Since $u'$ is strictly decreasing to zero,
$z\to\infty$ if and only if $W(z)\to\infty$.

Implicit differentiation gives
\[
    W'(z)=\frac{u'(W)}{A(W)}.
\]
Consequently,
\[
    \ell_0'(z)
    =u'(W)W'(z)
    =\frac{u'(W)^2}{A(W)}.
\]
Using $z=1/u'(W)$,
\[
    z\ell_0'(z)
    =\frac{u'(W)}{A(W)}
    =\left[\frac{A(W)}{u'(W)}\right]^{-1}.
\]
The ratio in brackets is positive, so Condition~\eqref{condition:U2} is equivalent to
$\lim_{z\to\infty}z\ell_0'(z)=0$.
\end{proof}

\subsection{Proof of Proposition~\ref{prop:relaxed-optimal-contract}}
\label{subsec:appendix-relaxed}

This section proves Proposition~\ref{prop:relaxed-optimal-contract}. This is a standard argument made in similar models \parencite{holmstrom1978incentives,jewitt2008moral}. We include it for completeness.

\begin{lemma}[Pointwise minimization]
\label{lem:lagrangian}
For $\lambda\geq0$ and $\mu\in\mathbb R$, the almost everywhere unique
minimizer of the lagrangian $\widehat{\mathcal L}(v,\lambda,\mu)$ over $\mathcal V$ is
$v_{\lambda,\mu}$.
\end{lemma}

\begin{proof}
Up to terms independent of $v$, the Lagrangian is the integral with measure
$f(y\mid a_0)d\nu(y)$ of
\[
    k(v(y))-w_0-[\lambda+\mu s(y)]v(y).
\]
This expression is strictly convex in $v(y)$.  Its interior first-order
condition is $k'(v(y))=\lambda+\mu s(y)$.  Taking limited liability into account,
the solution is $v(y)=\ell(\lambda+\mu s(y))$.
The assumption that canonical contracts are feasible ensures that this solution belongs to $\mathcal V$.  Strict convexity gives almost
everywhere uniqueness.
\end{proof}

\begin{lemma}[The local incentive compatibility multiplier]
\label{lem:mu-tilde}
For each $\lambda\geq0$, there is a unique $\widetilde\mu(\lambda)>0$ such
that the canonical contract $v_{\lambda,\widetilde\mu(\lambda)}$ satisfies
equation~\eqref{LIC_hat}.  Moreover, $\widetilde\mu$ is continuous.
\end{lemma}

\begin{proof}
For the proof, define
\[
    I(\lambda,\mu)
    :=\int \ell(\lambda+\mu s(y))s(y)f(y\mid a_0)\,d\nu(y).
\]
Because the score has mean zero, this can equivalently be written as
\begin{equation}
 I(\lambda,\mu)
 =\int[\ell(\lambda+\mu s(y))-\ell(\lambda)]s(y)
 f(y\mid a_0)\,d\nu(y).
 \label{eq:I-centered}
\end{equation}
Equation~\eqref{LIC_hat} is the same as $I(\lambda,\mu)=c'(a_0)$.

Equation~\eqref{eq:I-centered} shows that $I(\lambda,0)=0$ and that
$I(\lambda,\mu)$ is weakly increasing in $\mu$ because the term in brackets has the same sign as the score.  As $\mu\to\infty$,
monotone convergence yields
\[
 I(\lambda,\mu)\longrightarrow
 \Delta u\int_{\{s>0\}}s(y)f(y\mid a_0)\,d\nu(y).
\]
This limit exceeds $c'(a_0)$ by
\eqref{eq:local-implementability}.  Continuity therefore gives a positive
solution.

Whenever $I(\lambda,\mu)>0$, it is strictly increasing in $\mu$.
Therefore the multiplier satisfying equation~\eqref{LIC_hat} is unique.  Finally, joint continuity
of $I$ and this strict crossing imply continuity of
$\widetilde\mu(\lambda)$.
\end{proof}

For each $\lambda\geq0$, define the canonical contract satisfying equation~\eqref{LIC_hat} and its attained utility and cost by
\begin{eqnarray*}
    v_\lambda&:=&v_{\lambda,\widetilde\mu(\lambda)},\\
    \widetilde U(\lambda)&:=&\widehat U(v_\lambda,a_0),\\
    \widetilde C(\lambda)&:=&\widehat C(v_\lambda,a_0).
\end{eqnarray*}

The canonical contracts solving local incentive compatibility therefore form a one-dimensional path indexed by
$\lambda$.  The next lemma shows that these contracts solve a Pareto problem indexed by $\lambda$.

\begin{lemma}[Canonical contracts solve a Pareto problem]
\label{lem:pareto-problem}
For every $\lambda\geq0$, $v_\lambda$ is the almost everywhere unique
minimizer of
\[
    \widehat C(v,a_0)-\lambda\widehat U(v,a_0)
\]
among contracts in $\mathcal V$ satisfying equation~\eqref{LIC_hat}.  The function
$\widetilde U$ is
continuous and nondecreasing.  If $\lambda_1<\lambda_2$ and
$\widetilde U(\lambda_1)=\widetilde U(\lambda_2)$, then
$v_{\lambda_1}=v_{\lambda_2}$ almost everywhere.
\end{lemma}

\begin{proof}
Under equation~\eqref{LIC_hat}, the displayed objective differs by a constant from the
Lagrangian with $\mu=\widetilde\mu(\lambda)$.  Lemma~\ref{lem:lagrangian} therefore proves optimality and uniqueness.

For $\lambda_1<\lambda_2$, optimality of $v_{\lambda_1}$ at
$\lambda_1$ and of $v_{\lambda_2}$ at $\lambda_2$ gives
\begin{align*}
    \widetilde C(\lambda_1)-\lambda_1\widetilde U(\lambda_1)
    &\leq
    \widetilde C(\lambda_2)-\lambda_1\widetilde U(\lambda_2),\\
    \widetilde C(\lambda_2)-\lambda_2\widetilde U(\lambda_2)
    &\leq
    \widetilde C(\lambda_1)-\lambda_2\widetilde U(\lambda_1).
\end{align*}
Adding these inequalities gives
\[
    (\lambda_2-\lambda_1)
    [\widetilde U(\lambda_2)-\widetilde U(\lambda_1)]\geq0.
\]
If the utility levels are equal, both contracts minimize the same strictly
convex Pareto problem and hence agree almost everywhere.  Continuity follows
from Lemma~\ref{lem:mu-tilde} and canonical-contract regularity.
\end{proof}

Define
\[
    U_L:=\widetilde U(0).
\]
Recall that $U_R$ is the supremum feasible reservation utility for the relaxed
problem.  The next lemma shows that the Pareto path converges to this frontier
and rules out a terminal flat portion.

\begin{lemma}[The right utility frontier]
\label{lem:no-terminal-flat}
We have
\[
    \lim_{\lambda\to\infty}\widetilde U(\lambda)=U_R.
\]
Moreover, for every finite $M$, $\widetilde U(M)<U_R$.
\end{lemma}

\begin{proof}
Let
\[
    U_\infty:=\lim_{\lambda\to\infty}\widetilde U(\lambda),
\]
which is well defined in $\mathbb R\cup\{\infty\}$ by monotonicity.  Since
$v_\lambda$ is feasible for reservation utility $\widetilde U(\lambda)$, we
have $\widetilde U(\lambda)\leq U_R$ for every $\lambda$, and hence
$U_\infty\leq U_R$.

Conversely, for any contract $v$ satisfying equation~\eqref{LIC_hat}, Lemma~\ref{lem:pareto-problem} gives, for every $\lambda>0$,
\[
    \widehat U(v,a_0)
    \leq \widetilde U(\lambda)
       +\frac{\widehat C(v,a_0)-\widetilde C(\lambda)}{\lambda}
    \leq \widetilde U(\lambda)
       +\frac{\widehat C(v,a_0)}{\lambda}.
\]
Letting $\lambda\to\infty$ gives
$\widehat U(v,a_0)\leq U_\infty$.  Thus no reservation utility above
$U_\infty$ is feasible, so $U_R\leq U_\infty$.  Therefore
$U_\infty=U_R$.

Suppose that $\widetilde U(M)=U_R$ for some finite $M$.  Monotonicity makes
$\widetilde U$ constant on $[M,\infty)$, so Lemma~\ref{lem:pareto-problem} implies that every $v_\lambda$, $\lambda\geq M$,
agrees almost everywhere with a fixed contract $v$.  Local implementability
implies that $\{s(y)>0\}$ has positive measure.  On this set,
\[
    v(y)
    =\ell\bigl(\lambda+\widetilde\mu(\lambda)s(y)\bigr)
    \geq\ell(\lambda)
\]
almost everywhere.  Letting $\lambda\to\infty$ gives
$v(y)\geq\bar u$, a contradiction with $v$ being feasible.
\end{proof}

\begin{proof}[Proof of Proposition~\ref{prop:relaxed-optimal-contract}]
By \eqref{eq:utility-payment-equivalence}, the relaxed problem in payment
contracts is equivalent to the transformed problem in utility contracts.  We
prove the result for its utility-contract solution $v^*_{\bar U}$; applying
$w(y)=k(v(y))-w_0$ then gives the stated payment-contract solution.  By the
definition of $U_L$ and Lemma~\ref{lem:no-terminal-flat} with $M=0$, we have
$U_L<U_R$.

\emph{Part (i).}  Fix $\bar U<U_R$.  Continuity of $\widetilde U$ and
$\widetilde U(\lambda)\to U_R$ ensure that there is a $\lambda\geq0$ such that
\begin{equation}
    \widetilde U(\lambda)\geq\bar U,
    \label{eq:candidate-ir}
\end{equation}
and
\begin{equation}
    \lambda[\widetilde U(\lambda)-\bar U]=0.
    \label{eq:candidate-complementary-slackness}
\end{equation}
Set $\mu=\widetilde\mu(\lambda)$ and
$v^*_{\bar U}=v_\lambda$.  By construction, $v^*_{\bar U}$ is canonical and
satisfies equation~\eqref{LIC_hat} and IR.  By Lemma~\ref{lem:lagrangian}, $v^*_{\bar U}$ minimizes the Lagrangian (almost everywhere uniquely) and satisfies complementary slackness equation~\eqref{eq:candidate-complementary-slackness}. The
standard Lagrangian sufficiency argument implies that $v^*_{\bar U}$ is the almost
everywhere unique solution of the relaxed problem.

\emph{Part (ii).}  If $\bar U\leq U_L$, then $\lambda=0$ satisfies
\eqref{eq:candidate-ir} and
\eqref{eq:candidate-complementary-slackness}, so part (i) gives
$v^*_{\bar U}=v_0$.  Thus the relaxed-optimal contract is constant on this
region.  If $\bar U\in(U_L,U_R)$, the multiplier constructed in part (i) is
positive, so complementary slackness implies that IR binds.

On $(U_L,U_R)$, the set of associated IR multipliers is
\[
    \Lambda(\bar U)
    =\widetilde U^{-1}(\bar U)
    =\{\lambda\geq0:\widetilde U(\lambda)=\bar U\}.
\]
To see this, suppose first that $\lambda$ is an IR multiplier associated with
the solution.  Lemmas~\ref{lem:lagrangian} and \ref{lem:mu-tilde} imply that
the solution is $v_\lambda$.  Binding IR gives
$\widetilde U(\lambda)=\bar U$.  Conversely, if
$\widetilde U(\lambda)=\bar U$, then part (i) shows that $v_\lambda$ is the
relaxed solution with IR multiplier $\lambda$.

Lemma~\ref{lem:pareto-problem} shows that $\widetilde U$ is continuous and
nondecreasing.  Together with part (i) and
$\widetilde U(\lambda)\to U_R$, this implies that each $\Lambda(\bar U)$ is a
nonempty compact interval and that the multipliers are strictly ordered by
reservation utility as stated in the proposition.

Finally, fix any finite $M$.  Lemma~\ref{lem:no-terminal-flat} gives
$\widetilde U(M)<U_R$.  Hence, whenever
$\bar U>\widetilde U(M)$, every element of
$\widetilde U^{-1}(\bar U)$ exceeds $M$.  Therefore the smallest multiplier associated with $\bar U$ diverges to
infinity as $\bar U\uparrow U_R$.

\emph{Part (iii).}  The value $C^*$ is constant below $U_L$ by part (ii).
On $(U_L,U_R)$, the feasible sets are nested in reservation utility, so
$C^*$ is weakly increasing.  If $T_1<T_2$ and the two optimal costs were
equal, the optimal contract at $T_2$ would also solve the problem at $T_1$.
Almost everywhere uniqueness would then make the two optimal contracts equal,
contradicting binding IR.  Hence $C^*$ is strictly increasing.

Finally, consider two distinct reservation utilities in $(U_L,U_R)$.  Because
IR binds, their optimal contracts are distinct.  A strict convex combination
of these contracts satisfies local incentive compatibility and delivers the corresponding convex
combination of reservation utilities.  Strict convexity of expected cost pointwise in $y$
therefore implies that $C^*$ is strictly convex on $(U_L,U_R)$.
\end{proof}

\subsection{Proof of Proposition~\ref{prop:concave}}
\label{subsec:appendix-safe-tail}

This section proves Proposition~\ref{prop:concave}.  By
\eqref{eq:utility-payment-equivalence},
\[
    \widehat U_{aa}(v_{\lambda,\mu},a)
    =U_{aa}(w_{\lambda,\mu},a).
\]
Moreover, Lemma~\ref{lem:mu-tilde} implies that every canonical contract
satisfying equation~\eqref{LIC_hat} at a given $\lambda$ is $v_\lambda$.  It therefore suffices to
show that $\widehat U_{aa}(v_\lambda,a)\leq0$ for every $a$ when $\lambda$ is
sufficiently large. We begin with notation and preliminary results.

For any cutoff $q\leq0$, define
\begin{align}
    \Capq(q)&:=-\int_{\{s(y)\leq q\}}s(y)f(y\mid a_0)\,d\nu(y),
    \label{eq:cutoff-capacity}\\
    \Curvq(q)&:=\sup_{a\in\mathcal A}
    \int_{\{s(y)>q\}}[s(y)f_{aa}(y\mid a)]_+\,d\nu(y).
    \label{eq:cutoff-curvature}
\end{align}
$\Capq(q)$ is weakly increasing in $q$, while
$\Curvq(q)$ is weakly decreasing.
Moreover,
\begin{align}
 \sup_{q\in\operatorname{int}(\mathcal Q_{\mathrm{safe}})}\Capq(q)
 &=\mathrm{Cap}_{\mathrm{safe}},
 \label{eq:cutoff-capacity-limit}\\
 \inf_{q\in\mathcal Q_{\mathrm{safe}}}\Curvq(q)
 &=\mathrm{Curv}_{\mathrm{safe}}.
 \label{eq:cutoff-curvature-limit}
\end{align}

For $\lambda>m_0$, define the \textbf{kink cutoff} by
\[
    q^{\mathrm{kink}}(\lambda)
    :=\frac{m_0-\lambda}{\widetilde\mu(\lambda)}<0.
\]
Thus, limited liability binds exactly when
$s(y)\leq q^{\mathrm{kink}}(\lambda)$.  For any trial cutoff $q<0$, define
\begin{equation}
    \mu_q(\lambda):=\frac{m_0-\lambda}{q}>0.
    \label{eq:trial-multiplier}
\end{equation}
This is the multiplier that places the kink cutoff exactly at $q$.

\begin{lemma}[Kink placement]
\label{lem:kink-placement}
Fix $q<0$.  If
\[
    \Delta u\,\Capq(q)>c'(a_0),
\]
then, for all sufficiently large $\lambda$, the kink cutoff is weakly below
$q$:
\[
    q^{\mathrm{kink}}(\lambda)\leq q.
\]
Equivalently,
\[
    \widetilde\mu(\lambda)\leq\mu_q(\lambda).
\]
\end{lemma}

\begin{proof}
For $\lambda>m_0$, evaluate the left-hand side of equation~\eqref{LIC_hat} at the trial multiplier
$\mu_q(\lambda)$:
\begin{align*}
 &\int \ell\bigl(\lambda+\mu_q(\lambda)s(y)\bigr)s(y)
 f(y\mid a_0)\,d\nu(y)\\
 &=\int\left[\ell\bigl(\lambda+\mu_q(\lambda)s(y)\bigr)
              -\ell(\lambda)\right]s(y)f(y\mid a_0)\,d\nu(y)\\
 &\geq\int_{\{s(y)\leq q\}}
 \left[\ell\bigl(\lambda+\mu_q(\lambda)s(y)\bigr)
              -\ell(\lambda)\right]s(y)f(y\mid a_0)\,d\nu(y)\\
 &=\bigl[\ell(\lambda)-\underline u\bigr]\Capq(q).
\end{align*}
The first equality uses the mean-zero property of the score.  The inequality
follows because $\ell$ is nondecreasing: the centered difference has the same
weak sign as $s(y)$, so the centered integrand is nonnegative.  On $\{s(y)\leq q\}$,
$\lambda+\mu_q(\lambda)s(y)\leq m_0$, and hence
$\ell(\lambda+\mu_q(\lambda)s(y))=\underline u$; the final equality then
follows from the definition of $\Capq(q)$.

As $\lambda\to\infty$, the final expression converges to
$\Delta u\,\Capq(q)>c'(a_0)$.  Thus, for all sufficiently large $\lambda$, the
left-hand side of equation~\eqref{LIC_hat} evaluated at $\mu_q(\lambda)$
exceeds $c'(a_0)$.  Monotonicity in $\mu$ and uniqueness of the local
incentive compatibility multiplier therefore imply
$\widetilde\mu(\lambda)\leq\mu_q(\lambda)$.  Because $m_0-\lambda<0$, this
is equivalent to $q^{\mathrm{kink}}(\lambda)\leq q$.
\end{proof}

Kink placement and the curvature argument combine in the following
finite-$\lambda$ bound and its asymptotic consequence.

\begin{proposition}[Safe-tail curvature bound]
\label{prop:finite-lambda-bound}
Suppose $\ell_0$ is concave.  Let $q_0<q_1\leq0$ be safe cutoffs such that
\[
    \Delta u\,\Capq(q_0)>c'(a_0),
    \qquad \Curvq(q_1)<\infty.
\]
For all sufficiently large $\lambda$,
\begin{equation}
 \sup_{a\in\mathcal A}\widehat U_{aa}(v_\lambda,a)
 \leq
 \mu_{q_0}(\lambda)
 \ell_0'\bigl(\lambda+\mu_{q_0}(\lambda)q_1\bigr)\Curvq(q_1)
 -\underline c''.
 \label{eq:finite-lambda-bound}
\end{equation}
Moreover, if
\[
    \lim_{z\to\infty}z\ell_0'(z)=\rho<\infty,
\]
then
\begin{equation}
 \limsup_{\lambda\to\infty}\sup_{a\in\mathcal A}
 \widehat U_{aa}(v_\lambda,a)
 \leq
 \frac{\rho\,\Curvq(q_1)}{q_1-q_0}-\underline c''.
 \label{eq:asymptotic-safe-tail-bound}
\end{equation}
\end{proposition}

\begin{proof}
\mbox{}\par
\noindent\textbf{Step 1: The kink is eventually left of $q_0$.}\par
Lemma~\ref{lem:kink-placement} gives, for all sufficiently large $\lambda$,
\[
    q^{\mathrm{kink}}(\lambda)\leq q_0<q_1,
    \qquad
    \widetilde\mu(\lambda)\leq\mu_{q_0}(\lambda).
\]
Fix such a $\lambda$, abbreviate
$\widetilde\mu=\widetilde\mu(\lambda)$, and write
\[
    \delta_\lambda(t):=
    \ell(\lambda+\widetilde\mu t)-\ell(\lambda),
    \qquad t\in\mathbb R.
\]

\noindent\textbf{Step 2: center and split the curvature integral.}\par
We have
\begin{equation}
 \widehat U_{aa}(v_\lambda,a)
 =\int \delta_\lambda(s(y))f_{aa}(y\mid a)\,d\nu(y)-c''(a).
 \label{eq:curvature-identity}
\end{equation}
because the integral of $f_{aa}$ is zero, so that we can add and subtract
$\ell(\lambda)$ without changing the integral.

On $\{s(y) \leq q_1\}$, monotonicity of $\ell$ gives
$\delta_\lambda(s(y))\leq0$, while tail safety gives $f_{aa}\geq0$.  This
region therefore contributes weakly negatively. Therefore,
\begin{equation}
 \widehat U_{aa}(v_\lambda,a)
 \leq\int_{\{s(y)>q_1\}}
 \delta_\lambda(s(y))f_{aa}(y\mid a)\,d\nu(y)-c''(a).
 \label{eq:curvature-right-region}
\end{equation}

\noindent\textbf{Step 3: bound the right region.}\par

Kink placement implies
$\lambda+\widetilde\mu q_1>m_0$.  Since $\ell_0$ is increasing and concave,
for all $t>q_1$,
\[
    0\leq\delta_\lambda'(t)
    =\widetilde\mu\,\ell_0'(\lambda+\widetilde\mu t)
    \leq\widetilde\mu\,\ell_0'(\lambda+\widetilde\mu q_1).
\]
Since $\delta_\lambda(0)=0$, the mean-value theorem gives, on
$\{s>q_1\}$,
\[
    \delta_\lambda(s(y))f_{aa}(y\mid a)
    \leq
    \widetilde\mu\,\ell_0'(\lambda+\widetilde\mu q_1)
    [s(y)f_{aa}(y\mid a)]_+.
\]
Substituting this bound into \eqref{eq:curvature-right-region} and using the
definition of $\Curvq(q_1)$ gives
\begin{equation}
 \widehat U_{aa}(v_\lambda,a)
 \leq
 \widetilde\mu\,\ell_0'(\lambda+\widetilde\mu q_1)\Curvq(q_1)-c''(a).
 \label{eq:curvature-split}
\end{equation}

\noindent\textbf{Step 4: use the trial multiplier.}\par
Because $q_1\leq0$ and $\ell_0'$ is nonincreasing, $\ell_0'(\lambda+\mu q_1)$ is nondecreasing in $\mu$.  Therefore, $\mu\ell_0'(\lambda+\mu q_1)$ is nondecreasing in $\mu$ on the relevant nonbinding range.  Hence, replacing
$\widetilde\mu$ in \eqref{eq:curvature-split} by $\mu_{q_0}(\lambda)$ and
using $c''(a)\geq\underline c''$ proves
\eqref{eq:finite-lambda-bound}.

\noindent\textbf{Step 5: take the limit.}\par
Since $\lambda+\mu_{q_0}(\lambda)q_0=m_0$, the coefficient in
\eqref{eq:finite-lambda-bound} can be written, omitting the dependence of $\mu_{q_0}$ on $\lambda$, as
\begin{align*}
 &\mu_{q_0}
 \ell_0'\bigl(\lambda+\mu_{q_0}q_1\bigr)\\
 &\quad=
 \frac{\mu_{q_0}}
 {m_0+\mu_{q_0}(q_1-q_0)}
 \Bigl[m_0+\mu_{q_0}(q_1-q_0)\Bigr]
 \ell_0'\Bigl(m_0+\mu_{q_0}(q_1-q_0)\Bigr).
\end{align*}
Because $q_0<0$, $\mu_{q_0}(\lambda)\to\infty$ by equation~\eqref{eq:trial-multiplier}.  The fraction therefore
converges to $1/(q_1-q_0)$.  The product in brackets with $\ell_0'$
converges to $\rho$ by the assumption in the proposition statement.  Taking the limsup in \eqref{eq:finite-lambda-bound} proves
\eqref{eq:asymptotic-safe-tail-bound}.
\end{proof}

We can now prove Proposition~\ref{prop:concave}.

\begin{proof}[Proof of Proposition~\ref{prop:concave}]
By the equivalence established at the beginning of this subsection, it remains
to apply Proposition~\ref{prop:finite-lambda-bound} under each branch of
Theorem~\ref{thm:main}.

\emph{Branch (i): sufficient risk aversion.}
Condition~\eqref{condition:U1} and Remark~\ref{rem:link-concavity} imply that
$\ell_0$ is concave.  The capacity assumption
\[
    \Delta u\,\mathrm{Cap}_{\mathrm{safe}}>c'(a_0),
\]
together with \eqref{eq:cutoff-capacity-limit}, provides
$q_0\in\operatorname{int}(\mathcal Q_{\mathrm{safe}})$ such that
\[
    \Delta u\,\Capq(q_0)>c'(a_0).
\]
In particular, $q_0<0$.  The assumption
$\mathrm{Curv}_{\mathrm{safe}}<\infty$, equation~\eqref{eq:cutoff-curvature-limit}, and monotonicity of $\Curvq$ then allow us
to choose a safe cutoff $q_1\in(q_0,0]$ with $\Curvq(q_1)<\infty$.
Condition~\eqref{condition:U2} and Remark~\ref{rem:link-flattening} give
$z\ell_0'(z)\to0$.  Applying equation~\eqref{eq:asymptotic-safe-tail-bound} from Proposition~\ref{prop:finite-lambda-bound} with $\rho=0$ yields
\[
 \limsup_{\lambda\to\infty}\sup_{a\in\mathcal A}
 \widehat U_{aa}(v_\lambda,a)
 \leq-\underline c''<0.
\]

\emph{Branch (ii): log utility.}
For $u(W)=\log W$, we have $\ell_0(z)=\log z$, so $\ell_0$ is
concave, $z\ell_0'(z)=1$, and $\Delta u=\infty$.
Let
\[
    \underline s:=\operatorname*{ess\,inf}_{y\in\mathcal Y}s(y).
\]
The set $\mathcal Q_{\mathrm{safe}}$ is downward closed and hence is an
interval.  The definitions \eqref{eq:safe-curvature} and
\eqref{eq:safe-width} and the monotonicity of $\Curvq$ then imply
\begin{equation}
    \inf_{\substack{q_0,q_1\in\mathcal Q_{\mathrm{safe}}\\
                    \underline s<q_0<q_1}}
    \frac{\Curvq(q_1)}{q_1-q_0}
    =
    \frac{\mathrm{Curv}_{\mathrm{safe}}}
         {\mathrm{Width}_{\mathrm{safe}}},
    \label{eq:cutoff-ratio-limit}
\end{equation}
with the conventions following \eqref{eq:safe-width} when an endpoint
is infinite.  The strict inequality
\eqref{eq:log-curvature-width-condition} in Theorem~\ref{thm:main} therefore
provides safe cutoffs $\underline s<q_0<q_1\leq0$ such that
\begin{equation}
    \frac{\Curvq(q_1)}{q_1-q_0}<\underline c''.
    \label{eq:log-cutoff-ratio}
\end{equation}

Because $\underline s<q_0<0$, the set $\{s<q_0\}$ has positive probability
under $a_0$.  Hence \eqref{eq:cutoff-capacity} gives
\[
    \Capq(q_0)
    \geq-\int_{\{s<q_0\}}s(y)f(y\mid a_0)\,d\nu(y)>0.
\]
Since $\Delta u=\infty$, it follows that
$\Delta u\,\Capq(q_0)>c'(a_0)$.  Thus Proposition~\ref{prop:finite-lambda-bound} applies, and its bound with $\rho=1$ yields
\[
 \limsup_{\lambda\to\infty}\sup_{a\in\mathcal A}
 \widehat U_{aa}(v_\lambda,a)
 \leq\frac{\Curvq(q_1)}{q_1-q_0}-\underline c''<0.
\]

Thus, under either branch, $\widehat U_{aa}(v_\lambda,a)\leq0$ for every
$a\in\mathcal A$ and all sufficiently large $\lambda$.  Equation~\eqref{eq:utility-payment-equivalence} proves the proposition in its stated
payment-contract form.
\end{proof}

\end{document}